\documentclass[twocolumn]{aastex631}

\usepackage{xspace}

\usepackage{float}
\usepackage{enumitem}
\usepackage{amsmath}
\usepackage{longtable}

\graphicspath{{./}{.}}

\newcommand{\pynbody}{\texttt{pynbody}\xspace}
\newcommand{\lmstar}{$\log(M_{\star}/M_{\odot})$\xspace}
\newcommand{\lmass}{$\log(M/M_{\odot})$\xspace}
\newcommand{\lmsol}{$\log M_{\odot}$\xspace}
\newcommand{\msol}{$M_{\odot}$\xspace}
\newcommand{\zzero}{z=$0$\xspace}
\newcommand{\galaxyname}[1]{\texttt{#1}\xspace}
\newcommand{\HII}{${\rm H}_{2}$\xspace}

\usepackage[dvipsnames]{xcolor}
\definecolor{gedablue}{RGB}{48,97,243}
\definecolor{gedared}{RGB}{255,0,0}
\definecolor{gedaorange}{RGB}{255,69,0}

\newcommand{\extendedcolor}[1]{\textcolor{gedablue}{#1}}
\newcommand{\compactcolor}[1]{\textcolor{Red}{#1}}
\newcommand{\diskcolor}[1]{\textcolor{gedaorange}{#1}}
\newcommand{\nodiskcolor}[1]{#1}

\begin{document}
\title{The Formation of Dwarf Galaxy Disks}

\author[0000-0003-1509-9966]{Robel Geda}
\affiliation{Department of Astrophysical Sciences, Princeton University, 4 Ivy Lane, Princeton, NJ 08544, USA}
\email{robel@princeton.edu}

\author[0000-0001-7831-4892]{Akaxia Cruz}
\affiliation{Center for Computational Astrophysics, Flatiron Institute, New York, NY 10010, USA}
\affiliation{Department of Physics, Princeton University, Princeton, NJ 08544, USA}
\affiliation{Department of Astrophysical Sciences, Princeton University, 4 Ivy Lane, Princeton, NJ 08544, USA}

\author[0000-0002-1685-5818]{Anna C.\ Wright}
\affiliation{Center for Computational Astrophysics, Flatiron Institute, New York, NY 10010, USA}

\author[0000-0002-5612-3427]{Jenny E. Greene}
\affiliation{Department of Astrophysical Sciences, Princeton University, 4 Ivy Lane, Princeton, NJ 08544, USA}

\author[0000-0002-0372-3736]{Alyson Brooks}
\affiliation{Center for Computational Astrophysics, Flatiron Institute, New York, NY 10010, USA}
\affiliation{Department of Physics \& Astronomy, Rutgers, the State University of New Jersey, Piscataway, NJ, 08854, USA}

\author[0000-0001-5510-2803]{Thomas Quinn}
\affiliation{Astronomy Department, University of Washington, Box 351580, Seattle, WA, 98195-1580, USA}

\author[0000-0001-8745-0263]{James Wadsley}
\affiliation{Department of Physics and Astronomy, McMaster University, 1280 Main Street West, Hamilton, Ontario, L8S 4M1, Canada}

\author[0000-0002-9642-7193]{Ben Keller}
\affiliation{Department of Physics and Materials Science, University of Memphis, 3720 Alumni Avenue, Memphis, TN 38152, USA}

\begin{abstract}
Dwarf galaxies are dark matter-dominated systems that are sensitive to feedback and display a diversity of baryonic morphologies. This makes them excellent probes for understanding dark matter and galaxy evolution. This work investigates the physical processes that influence the sizes of isolated dwarf galaxies using high-resolution cosmological zoom-in simulations of $39$ dwarf galaxies drawn from the Marvelous Massive Dwarfs simulation suite ($7.5 <$ \lmstar $< 9.1$). Our simulations show that dwarf galaxies initially form as compact galaxies ($R_e < 2$ kpc). However, several of these galaxies ($54\%$) experience periods of gradual size growth at relatively stable sSFR, allowing them to become extended galaxies. While previous simulations have struggled to produce dwarf galaxy disks, we find that the growth of rotation-supported stellar disks is the primary means by which isolated dwarfs become extended in size. These stellar disks are formed by mergers with high orbital angular momentum satellites on high angular momentum (spiraling-in) orbits, which spin up the gas surrounding the central galaxy and contribute $\approx 30 \%$ of the cold gas mass at \zzero. For these systems, star formation in the angular momentum supported gas and the gradual build up of stars in the disk result in secular size growth. 
\end{abstract}

\accepted{April 2026}

\section{Introduction}
\label{sec:Introduction}
Dwarf galaxies serve as ideal laboratories for studying both dark matter and galaxy evolution. Dark matter tends to dominate their gravitational potential, so stellar and gas kinematics trace the underlying mass distribution with minimal bias \citep[e.g.,][]{1997PASP..109..745B, 2022NatAs...6..659B, 2022ApJ...940....8M, 2025MNRAS.541.2180R}. Furthermore, their shallow potentials amplify the effects of feedback and environmental processes, exhibiting a wide range of morphologies and density profiles \citep{1986ApJ...303...39D, 1999ApJ...513..142M, 2000MNRAS.313..291F, 2009ARA&A..47..371T, 2019ApJ...886...74M}. By studying dwarf galaxies, it is possible to test how dark matter and baryons interact on the smallest scales. 

Modern zoom–in simulations struggle to reproduce the remarkable variety of baryonic morphologies, with many lacking dwarf disk formation \citep{Benavides2025, Celiz2025}. Beyond this, they tend to lack rotation curve shape diversity, and observed inner density profiles that dwarf galaxies display \citep[e.g.,][]{2015MNRAS.452.3650O, 2018MNRAS.473.4392S, 2020MNRAS.495...58S}. Thus, dwarf galaxy sizes and rotation curves provide a critical and stringent test of galaxy formation models in simulations. Despite the diversity in size and rotation speed, galaxies empirically obey a size-mass relation \citep{2003MNRAS.343..978S, 2015MNRAS.447.2603L, 2022ApJ...933...47C}. The influence and relative importance of mergers, angular momentum (AM) transport, star formation, and feedback on this relation are not well understood \citep[e.g.,][]{2023MNRAS.518.1002R}. 

In addition to galaxy size, observational kinematic and morphological studies further provide important constraints on disk formation. Spatially resolved kinematic studies of dwarf galaxies show that galaxies exhibit increasing rotational support toward higher stellar mass \citep[e.g.,][]{2007ApJ...660L..35K, 2012ApJ...758..106K}. Similarly, observational morphological studies indicate that star forming galaxies span a wide range of intrinsic shapes, with measurable trends in diskiness and thickness as a function of mass  \citep{2019MNRAS.484.5170Z, 2020ApJ...900..163K, 2024ApJ...963...54P}. In particular, massive dwarf galaxies ($8.5 \lesssim$ \lmstar $\lesssim 9.5$) are interesting because they are natural testbeds for studies of the mass scale associated with disk formation \citep[e.g.,][]{2015MNRAS.452..986S, 2017MNRAS.465.2420W, 2023ApJ...951...52D, Keith2025, 2025MNRAS.541.2180R, Benavides2025}.

Dwarf galaxies have been observed to undergo cyclic star formation histories, referred to as ``bursty" star formation, in which their star formation fluctuates around a baseline formation rate \citep{2009ARA&A..47..371T, 2010ApJ...724...49M, 2010ApJ...721..297M, 2011ApJ...739....5W, 2014ApJ...789...96A, 2022MNRAS.511.4464A, 2023ApJ...950..125M}. Because of their low masses, dwarf galaxies simulations show supernova driven outflows from these episodes of intense star formation can expand the stellar half-mass radius, while also flattening the central density profiles \citep[e.g,][]{2009MNRAS.395.1455S, 2012MNRAS.421.3464P, 2012ApJ...755L..35M, 2013MNRAS.429.3068T,  2014MNRAS.437..415D, 2015MNRAS.454.2981C, 2015MNRAS.448..792G, ElBadry16, NIHAO2019, 2024ApJ...977...20R}. \cite{ElBadry16} in particular found a coupling between feedback and size that results in anti-correlation between sSFR and galaxy size (``breathing mode") for the FIRE dwarf galaxies. The cycle begins with the formation of new stars that generate feedback, pushing material away from the center of the galaxy. Over time, this expelled material returns towards the galaxy's center, sparking new star formation, which will initiate the next cycle. Observational tests of these predictions have resulted in mixed results \citep{patel18,2023ApJ...945...93P,2019ApJ...880...54H,2020ApJ...896L..26P,2021ApJ...922..217E, 2025arXiv251017996L}, with some observations indicating that the bursts of star formation may be too brief and intense in the FIRE simulations \citep[e.g.][]{2010ApJ...724...49M,2010ApJ...721..297M,2025arXiv250616510M}. \cite{patel18} in particular observe a sample of $284$ isolated intermediate mass galaxies ($9 < M_{\star} < 9.5$ \lmsol) within a redshift of $0.3 < z < 0.4$ to measure their size-sSFR distribution. They report opposing trends to the breathing modes predicted for the most extreme galaxies in the FIRE simulations. In addition, they report intermediate mass galaxies with lower sSFRs are relatively compact and have more significant bulge contributions to their light profiles.

In this work, we examine high spatial and mass resolution simulations of $39$ low-mass systems ($7.5 <$ \lmstar $< 9.1$). These galaxies lie in the regime where bursty star formation is expected, making them an ideal sample for this study. We follow their evolutionary histories, focusing on the connection between size and sSFR. Motivated by \cite{patel18}, we also examine the size-sSFR evolution of galaxies in our sample to determine the role feedback plays in the sizes of dwarf galaxies. Our study suggests a dichotomy where some dwarfs grow substantially in size at late times by forming and sustaining rotation-supported disks, while others remain compact throughout cosmic time.  Building on results that mergers torque disk AM in Milky Way mass halos \citep[e.g.,][]{2002ApJ...581..799V,Cox2006, 2021MNRAS.503.2866D, Cadiou2022, Simons2024, Joshi2025}, we interrogate what factors nurture or stymie disk formation and growth, with particular attention to the role of mergers in transferring AM and assembling extended stellar components. Furthermore, the relatively high spatial resolution of our simulations makes it possible to track angular momentum transfer reliably \citep{2004ApJ...607..688G, 2007MNRAS.375...53K}.

We outline our simulation, sample selection, and related measurements in Section \ref{sec:Methods}. We discuss the size-sSFR relation of our sample in Section \ref{sec:size-sSFR relation}. Section \ref{sec:Extended Galaxies and Stellar Disks} discusses how disk formation is related to size, and what role mergers play in the disk formation process. Section \ref{sec:Discussion} provides discussions on our results. Finally, Section \ref{sec:conclusion} provides a summary of this work and concluding remarks. Throughout this work, we report distances in physical units and use Planck 2015 cosmology \citep[$\Omega_{m}=0.3086$, $\Omega_{\Lambda}=0.6914$, $h = 0.67$, and $\sigma_{8}=0.77$,][]{2014A&A...571A..16P} unless otherwise stated.

\section{Methods}
\label{sec:Methods}

\begin{figure*} 
\includegraphics[width=\textwidth]{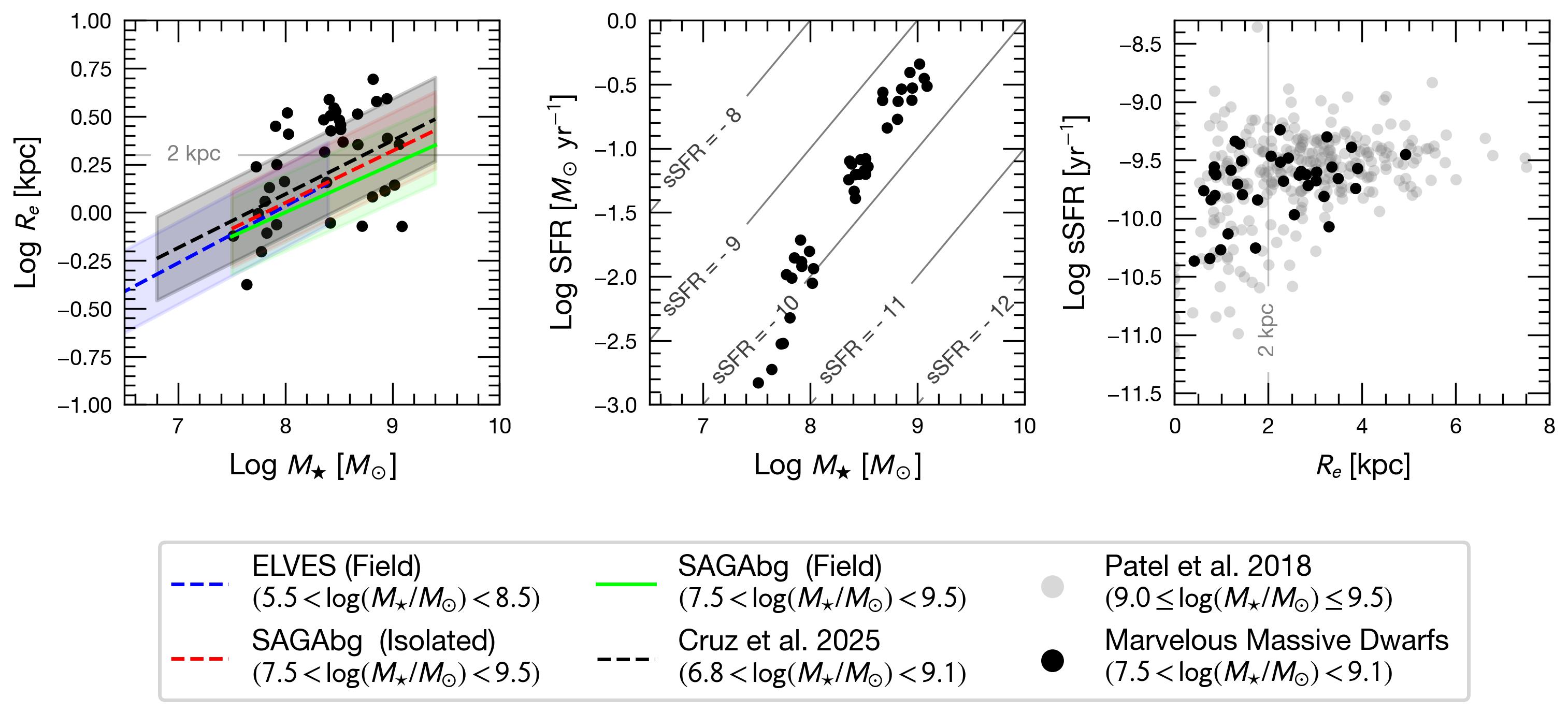}
\centering
\caption{\label{fig:z0_tri_panel}
The mass-size, mass-SFR, and size-sSFR relations of our final sample of galaxies at present day (\zzero, as black points). We add a $2$ kpc reference line in the mass-size and size-sSFR panels (further discussed in Section \ref{sub:Secular Size Growth}). In the first panel we include observational mass-size relations and $1\sigma$ errors from ELVES \citep{Carlsten2021} and SAGA \citep{SAGA} for reference. We also include the mass-size relation from \cite{Cruz2025}, who combine the Marvelous Massive Dwarf and Extended Dwarf Zoom simulations. In the size-sSFR panel, we include galaxies from \cite{patel18} (in gray) for reference. We note that the \cite{patel18} sample occupies a narrower and higher mass bin than our population of galaxies, thus is not ideal for comparison to our sample despite the apparent consistency.
}
\end{figure*}

\subsection{Simulations}
\label{sub:Simulation}
The Marvelous Massive Dwarfs \citep[referred to as ``Massive Dwarfs" throughout this paper simulation suite,][]{Cruz2025, Keith2025, 2025MNRAS.541.2180R, 2025ApJ...993..230P} consists of high-resolution cosmological zoom-in simulations of individual dwarf galaxies in a cosmological context \citep[e.g.,][]{1993ApJ...412..455K, 2008MNRAS.390.1349P, 2014MNRAS.437.1894O, Hopkins2014}. The initial conditions for these simulations were drawn from the \texttt{Romulus25} simulation volume \citep[$(25\ {\rm Mpc})^3$,][]{Tremmel2017}, with the zoom region itself much smaller ($\sim 300-400$ kpc). The simulations target isolated dwarf galaxies spanning stellar masses of approximately $7.5 \leq$ \lmstar $\leq 9.1$ ~at $z=0.1$. We note that despite selecting our sample at $z=0.1$, the \texttt{Romulus25} simulations were run to \zzero. We define isolation as having no neighboring galaxies with stellar masses \lmstar $\geq 10$ within $1.5$ Mpc of the central galaxy \citep{2012ApJ...757...85G}. These simulations were performed using the N-body+SPH code \texttt{ChaNGa} \citep{Jetley2008, Jetley2010, Menon2015}, which incorporates hydrodynamic modules from \texttt{GASOLINE2} \citep{Wadsley2004, Wadsley2017}. Massive Dwarfs adopts a Planck 2015 cosmology \citep[$\Omega_{m}=0.3086$, $\Omega_{\Lambda}=0.6914$, $h = 0.67$, and $\sigma_{8}=0.77$,][]{2014A&A...571A..16P}. 

The simulations achieve a spline gravitational force resolution (softening length) of $87$ pc. The initial particle masses for this suite are $3.3 \times 10^{3}$\msol for gas particles, $994$\msol for star particles, and $1.8 \times 10^{4}$\msol for dark matter particles within the zoom-in region. The simulation outputs (snapshots) are written at approximately $300$ Myr intervals. Each zoom-in simulation adheres to the \galaxyname{r123} naming convention, where the number is its halo id in \texttt{Romulus25} at $z=0.1$. 

The star formation threshold is defined to be when a gas particle reaches temperatures below $1000$ K and at number densities above $0.1\ m_h\ {\rm cm}^{-3}$ \citep{2012MNRAS.425.3058C}. For gas particles that meet this threshold the simulations incorporate a modified version of star formation probabilities from \cite{2006MNRAS.373.1074S} that includes a probabilistic \HII abundance driven star formation scheme described by \cite{2012MNRAS.425.3058C}:

\begin{equation}
    p_* = \frac{m_{\rm g}}{m_{\rm s}} \left(1 - {\rm exp}\left[- \frac{c^{*}_{0} X_{{\rm H}_2} \Delta t}{t_{\rm ff}} \right]\right),
\end{equation}

Where $p_*$ is the star formation probability, $m_{\rm g}$ is the particle's gas mass, $m_{\rm s}$ is the initial mass of the star particle that would form, $c^{*}_{0}$ is the star formation efficiency (set to $0.1$), $X_{{\rm H}_2}$ is the fraction of \HII, $\Delta t$ is the time span of the timestep, and $t_{\rm ff}$ is the local free fall time. Due to the \HII driven star formation, stars usually form in gas that is much denser than the density threshold ($n \gtrsim 100\ m_h\ {\rm cm}^{-3}$). Lastly, the simulations include gas cooling as described by \cite{2012MNRAS.425.3058C}, metal line cooling and the diffusion of metals as described by \cite{2010MNRAS.407.1581S}, and the model accounts for the non-equilibrium formation and destruction of molecular hydrogen. 

Supernova (SN) feedback in Massive Dwarfs is governed by the superbubble model described by \cite{2014MNRAS.442.3013K}, which uses multiphase gas, evaporation and thermal conduction. Each SN event deposits $10^{51}$ erg of energy in a brief multi-phase state. In this SN model, evaporation from thermal conduction acts to regulate the amount of hot gas \citep{2014MNRAS.442.3013K}. The hot and cold phases each carry a separate mass, energy, and non-equilibrium hydrogen and helium ionization fractions. Cooling rates for each phase are also then calculated separately, with a density set assuming pressure equilibrium between the two phases. Lastly, Massive Dwarfs implements massive black hole (BH) physics with boosted Bondi accretion \citep{Booth2009} and feedback models from \cite{Tremmel2015, Tremmel2017}. The simulations have lower seed masses ($10^{5}$ \msol) than \texttt{Romulus25} and use the physical state of the gas at high redshifts for seed formation instead of BH halo occupation functions \citep{Tremmel2015,Tremmel2017,2019MNRAS.482.2913B}.   

\subsection{Halo Finding and Centering}
\label{sub:Halo Finding and Centering}

We use AMIGA's Halo Finder \citep[AHF;][]{2001MNRAS.325..845K, 2004MNRAS.351..399G} to identify dark matter halos and all bound particles (baryonic and dark). For each identified halo, we calculate the center of mass (COM method) with \pynbody \footnote{\url{https://github.com/pynbody/pynbody}}, using all baryonic and dark matter particles. Once the center of mass is determined, we approximate the bulk velocity of the halo using the average velocities of all bound particles within $2$ kpc of the center (physical units). At each snapshot, \pynbody orients the galaxy in a ``face-on" configuration by calculating the total AM vector (see Section \ref{sub:Angular Momentum}) of all bound gas particles within a radius of $2$ kpc from the center. Unless stated otherwise, we center the simulation on the main halo of each snapshot using its center of mass (using the same method as above) and measure velocities relative to its bulk velocity. For unbinding and our total mass (sum of bound mass) measurements, we use the virial mass ($M_{\rm vir}$) as defined by \cite{1998ApJ...495...80B}. We calculate the total stellar mass of a galaxy by summing over all bound stars in its parent halo. To compare with observational stellar masses derived from photometric measurements, we correct the sum of the star particle masses by applying a factor of $0.6$ to all galaxy stellar masses as described in \cite{2013ApJ...766...56M}, who showed that photometrically inferred stellar masses recover approximately $60\%$ of the total stellar mass in star particles for galaxies in our mass scale.

\subsection{Merger Trees}
\label{sub:Merger Trees}

\begin{figure*} 
\includegraphics[width=\textwidth]{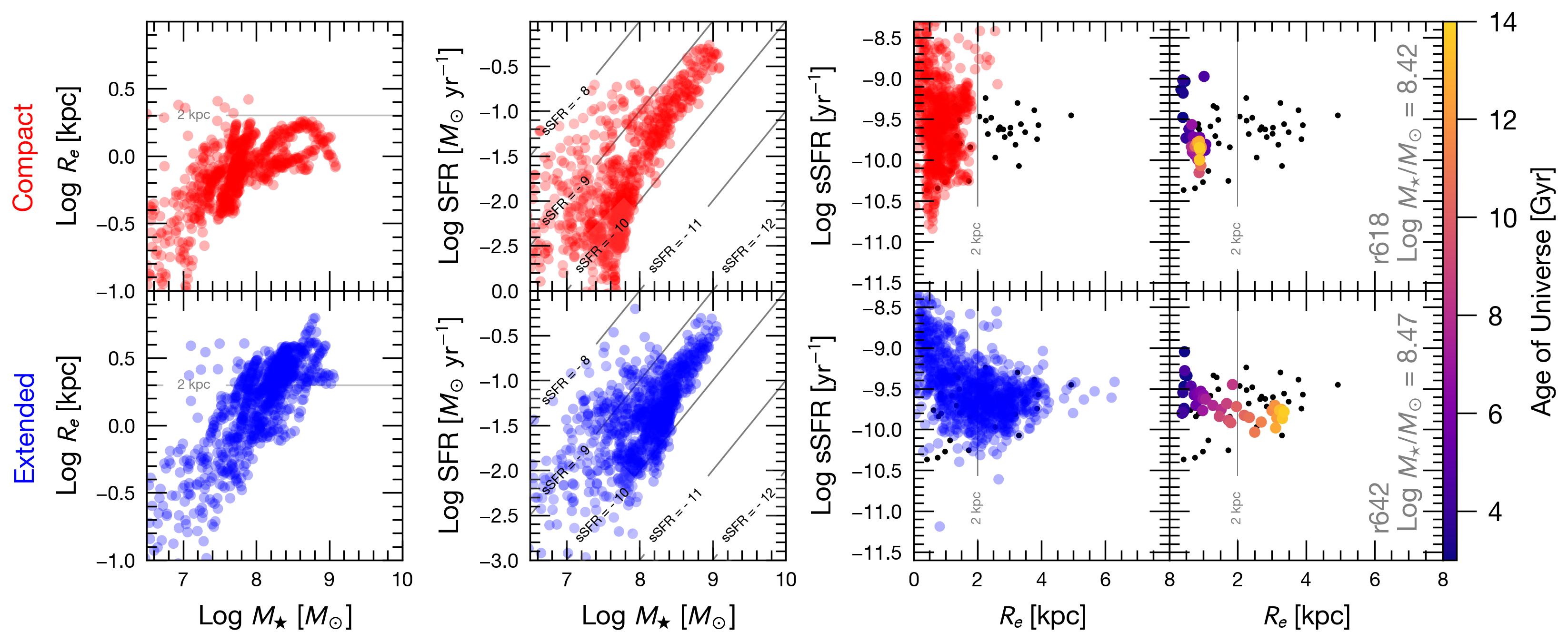}
\centering
\caption{\label{fig:four_panel_compact_vs_extended}
The evolution of galaxies in our compact (top row) and extended (bottom row) sub-samples as defined by \zzero half-light radii. Similar to Figure \ref{fig:z0_tri_panel}, the first three columns show the mass-size, mass-SFR, and size-sSFR relations for every galaxy in each sub-sample, at every snapshot up to $z=3$. The last column shows the evolution of a single galaxy example in each sub-sample across cosmic time. In all the size-sSFR panels, we show the \zzero values of all the galaxies in our final sample as black points for reference. In the compact sample, galaxies generally maintain a small size of less than $2$ kpc (with some spikes above $2$ kpc due to halo finder confusion). Extended dwarf galaxies begin as compact galaxies but eventually experience secular size growth over time at relatively stable sSFRs.
}
\end{figure*}

We link halos across different snapshots using their particle membership information. Briefly, for a main halo at a given snapshot, we identify which progenitor halo the majority of its dark matter particles belonged to in the previous snapshot. We refer to this progenitor as the main progenitor to distinguish it as contributing the most dark matter particles. We generate merger trees for each main halo up to $z=3$. 

We reconstruct the assembly history of the main halo by tracing all of its present-day (\zzero) particles backward in time. At each earlier snapshot, we identify all halos that host these particles and connect them to their descendants. We link sub-halos and non–main progenitors across snapshots using the “voter and graveyard” approach described in \cite{10.1093/mnras/staf016}. Voter particles are the most gravitationally bound particles (dark matter in this case) within each structure, which makes them most likely to be embedded within the same halo across snapshots \citep{2022MNRAS.510..959I}. In this work, voters are selected within a $800$ pc radius of the sub-halo's center to reduce contamination. At each snapshot, sub-halos with at least $1000$ particles are assigned $100$ voter particles. Any sub-halo in the subsequent snapshot that captures more than $40\%$ of these voter particles is identified as the voting sub-halo’s descendant. We find that this parameter combination (the number of voter particles and overlap threshold) produces sufficiently reliable connections for our simulation resolution.

In some instances, halos are temporarily undetectable (e.g., too close to the center of the main halo) and may reappear in a later snapshot. This disrupts the tracking process if we only consider descendant candidates in subsequent snapshots \citep[][]{2013ApJ...763...18B}. To account for this, we store and update the ID of the last halo each voter particle was a member of in a ``graveyard" array \citep{10.1093/mnras/staf016}. When a candidate progenitor fails to secure a descendant in the next snapshot, or if its descendant is the main halo, we index the graveyard array via each voter particle's ID and tally their most recent halo memberships. If $\geq 20\%$ of them point to the same non–main halo ID in a later snapshot, we link the candidate progenitor to that halo as its descendant. 

\subsection{Star Formation Rates}
\label{sub:Star Formation Rates}

The simulations record and output the precise formation times of each star particle, providing us with a very high temporal resolution ($\approx1$ Myr). With future comparisons to observational UV-based SFRs in mind \citep[e.g.,][]{patel18}, which trace star formation over $\sim100$ Myr timescales, we compute SFRs in $100$ Myr bins at each snapshot. sSFRs are then computed using the stellar mass of the galaxy (see Section \ref{sub:Halo Finding and Centering}) at the corresponding snapshot.

\subsection{Size Measurements}
\label{sub:Size Measurements}

\begin{figure} 
\includegraphics[width=8cm]{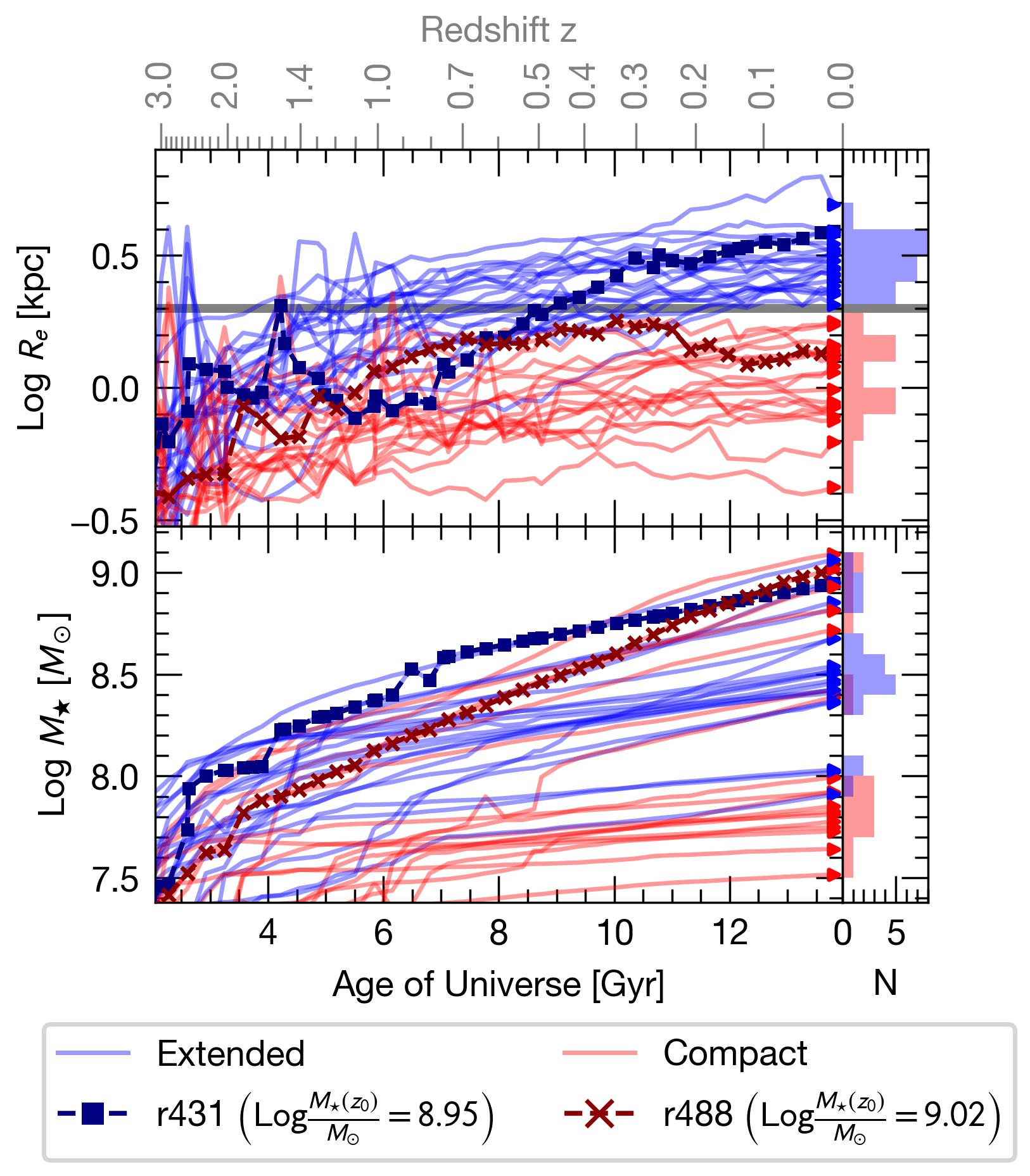}
\centering
\caption{\label{fig:r_and_m_hist}
The evolution of half-light radii (top) and stellar masses (bottom) of galaxies in our final sample. We show these quantities up to $z=3$. We do not filter out halo finder confusion in which mergers appear as spikes in size right before coalescence (usually appears before $6$ Gyr). Compact galaxies are shown in red and extended galaxies are in blue. We highlight an example galaxy (\lmstar $\approx 9.0$) for each category in both panels, with \galaxyname{r431} (extended) as blue squares and \galaxyname{r448} (compact) in red. We provide a histogram of both quantities at \zzero to the right of each panel. The compact and extended galaxies are separated in size by definition and cover a range of up to $4$ kpc. Though the two sub-samples are bifurcated by size, they cover a diverse range of masses. In particular, compact galaxies span both the low and high end of the stellar mass range. 
}
\end{figure}

In our analysis, we use the stellar half-light radius ($R_e$) as our primary size measure and create mock face-on observations of our galaxies by aligning the galaxy's gas AM vector with the z-axis (see Section \ref{sub:Halo Finding and Centering} and \ref{sub:Angular Momentum}). We use \pynbody to measure the half-light radius in the Johnson $R$ band (without extinction) in cylindrical coordinates (as seen from the $z$-axis). \pynbody first computes the total luminosity of a galaxy by treating each star particle as a simple stellar population (SSP) and assigning magnitudes interpolated from precomputed SSP tables \citep[generated using \texttt{CMD} from the Padova group;][]{2008A&A...482..883M, 2010ApJ...724.1030G}, based on the particle's age and metallicity. It then constructs a radial cumulative luminosity profile and identifies the half-light radius as the radius that encloses half of the total luminosity. Lastly, we note that the face-on size measurements are not fully representative of observations, which depend on viewing angle and surface brightness limits \citep{2022ApJ...926...92V}.

\subsection{Angular Momentum Measurements}
\label{sub:Angular Momentum Measurements}

\subsubsection{Angular Momentum}
\label{sub:Angular Momentum}
In this study, we probe the internal AM of the central galaxies, as well as the orbital angular momenta of the satellites. \pynbody computes the total AM of each main halo by summing the angular momenta of the particles within $R_{vir}$. We measure the specific AM of a component as follows:

\begin{equation}
    \vec{j_c}= \frac{\vec{L_c}}{M_c} = \frac{ \sum_{i}  \vec{r_i} \times m_i\vec{v_i}}{\sum_{i} m_i},
\end{equation}

where $\vec{j_c}$ is the specific AM vector of the component. $\vec{L_c}$ and $M_c$ are the AM vector and the total mass of the component, respectively. $m_i$ is the mass of each particle of the component, while $\vec{r}_i$ and $\vec{v}_i$ are the particle's position and velocity vectors relative to the main halo's center of mass and bulk velocity, respectively. We denote the magnitudes of the AM vectors as $j_c \equiv \left|\vec{j_c}\right|$.

\subsubsection{Disk Definition}
\label{sub:Disk Definition}
We quantify whether a galaxy has a significant stellar disk component using the specific AM distribution of its star particles. We first calculate the unit vector of the specific AM for each star particle ($\hat{j}$) and then construct a histogram for each Cartesian axis. From these histograms, we take the median value for each coordinate component. Since the galaxy is oriented face-on with respect to its total AM at every snapshot, the median of the $\hat{j_z}$ distribution will tend to be skewed towards $+1$ if the galaxy has a stellar disk; in contrast, the median values of $\hat{j_x}$ and $\hat{j_y}$ will be close to zero. We kinematically define a galaxy as having a disk if the median $\hat{j_z}$ of its star particles is greater than $0.5$.

\subsection{Simulated Galaxy Sample}
\label{sub:Galaxy Sample}

\begin{figure*} 
\includegraphics[width=\textwidth]{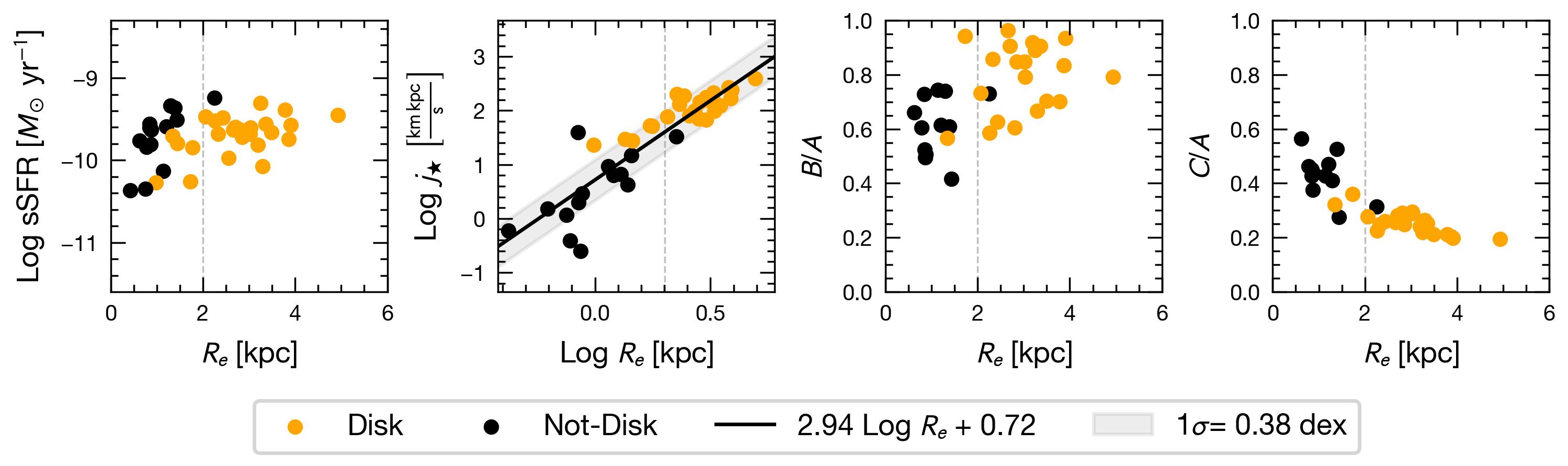}
\centering
\caption{\label{fig:disk_size_params}
Our disk and not-disk sub-samples. The first two panels from the left show the size-sSFR and size-$j_\star$ (where $j_\star \equiv |\vec{j_\star}|$) relations. The last two panels show stellar axis ratios of our sample measured at $R_e$  \citep[from][]{Keith2025} as a function of size ($A$ is the semi-major axis, $C$ is the semi-minor axis, and $B$ is the intermediate axis). Disk galaxies are shown in orange, while galaxies that are not-disks are in black. Almost all extended galaxies are disk galaxies, and disk galaxies tend to have higher stellar $j_\star$ since they are AM supported systems. 
}
\end{figure*}

Our simulation suite consists of a total of $41$ isolated dwarf galaxies. One of these galaxies, \galaxyname{r918}, is actively merging at redshift \zzero and we exclude it from further analysis. We exclude from our final sample a second galaxy, \galaxyname{r615}, which also recently experienced a major merger; however we discuss it as an individual case study in Section \ref{sub:Disrupted Disk}. Excluding the two merging galaxies, the final sample consists of $39$ galaxies with a total mass range of $10.2 \leq$ \lmass $\leq 11.0$ and stellar mass range of $7.5 <$ \lmstar $< 9.1$ (see Section \ref{sub:Halo Finding and Centering} for how mass is determined). A table of halo properties for our galaxy sample is presented in Table \ref{table:overview} in Appendix \ref{appendix:Table of Halo Properties}.

\section{Mass, Size, and Star Formation}
\label{sec:size-sSFR relation}

In this section, we examine the relationships between stellar mass, size ($R_e$), and SFR for our sample of galaxies. We present the \zzero mass-size, mass-SFR, and size-sSFR relations of our sample in Figure \ref{fig:z0_tri_panel}.
We first present the mass-size and mass-SFR relations, then focus on the size–sSFR relation to explore the coupling of galaxy size to star formation activity (i.e., breathing modes). Lastly, we examine the evolution of galaxy sizes as a function of cosmic time up to $z=3$.

\subsection{Mass-Size-SFR Relations at Redshift \zzero}
\label{sub:Mass-Size-sSFR Relations at Redshift}

In the first panel of Figure \ref{fig:z0_tri_panel} we present the mass-size relation for our sample of galaxies at present day. We expect that galaxy size is primarily a function of mass, with the two quantities being directly correlated. Details of the mass-size relation for our sample of galaxies is presented and discussed in \cite{Cruz2025} (see their Section 5.3). Their measurements indicate that the Marvelous Massive Dwarfs simulations agree better with SPARC \citep{SPARC} in the size-mass plane than other simulations, and with the size-mass relationship for  dwarfs as derived in \cite{Carlsten2021} \citep{Cruz2025}.\footnote{Note that while we measure sizes using a technique equivalent to curve of growth or Petrosian radii, and \cite{Carlsten2021} and \cite{SAGA} adopt Sérsic radii, there is nonetheless good agreement between these relations. We also find consistency with the SPARC \citep{SPARC} mass--size relation measured with Petrosian radii \citep{Cruz2025} (see, e.g., their Fig.~7).} Furthermore, the Marvelous Massive Dwarfs simulation sample has been shown to reproduce realistic star formation rates across stellar masses, with specific star formation rates broadly consistent with observational constraints, though with a tendency toward elevated sSFR at \lmstar $> 8.0$ \citep{2025ApJ...993..230P}. 

\subsection{Size-sSFR Relations at Redshift \zzero}
\label{sub:Mass-Size-sSFR Relations at Redshift}
The size-sSFR relation is especially relevant for dwarf galaxies in our mass range because they are predicted to be sensitive to feedback due to their weak gravitational potentials and are observed to experience cyclic star formation histories \citep{2009ARA&A..47..371T, 2010ApJ...724...49M, 2010ApJ...721..297M, 2011ApJ...739....5W, 2014ApJ...789...96A, 2022MNRAS.511.4464A, 2023ApJ...950..125M}. \cite{ElBadry16} find that the sizes of dwarf galaxies in the FIRE simulations are inversely related to their sSFRs due to a coupling of size to feedback. This inverse relation appears as an anti-correlation in the size-sSFR relation. 

In the third panel of Figure \ref{fig:z0_tri_panel} we present the size-sSFR relation of our sample at \zzero. Though the \cite{patel18} sample represents galaxies at a higher mass range ($9 <$ \lmstar $< 9.5$) than our sample ($7.5 <$ \lmstar $< 9.1$), we include their observed size-sSFR relations for isolated intermediate mass galaxies in gray for reference. We note that the \zzero distribution of our sample does not show a strong anti-correlation at \zzero. Because breathing modes occur in quasi-periodic cycles, we further investigate the size-sSFR evolution of our galaxies as a function of time.

\subsection{Compact and Extended Galaxies}
\label{sub:Compact and Extended Galaxies}

To investigate the evolution of the size-sSFR relation for our sample, we examine the sizes and sSFRs of all our galaxies up to $z=3$ (the highest redshift in our merger trees). Our measurements, shown in Figure \ref{fig:four_panel_compact_vs_extended}, indicate that the values for our galaxies fall within the general \zzero region throughout cosmic time. More importantly, we find two populations of galaxies that can be distinguished using their half-light radii in the size-sSFR relation. Galaxies with a half-light radius of less than $2$ kpc at \zzero typically maintain sizes below $2$ kpc throughout their evolutionary timeline. This suggests that the size of such galaxies is a relatively consistent characteristic that remains stable as their sSFRs evolve over cosmic time. We refer to these galaxies as compact galaxies. In comparison, we observe that galaxies with a half-light radius greater than $2$ kpc at \zzero start compact at high redshifts but increase in size over cosmic time. We refer to these galaxies as extended galaxies. The compact sub-sample at \zzero comprises $18$ galaxies with stellar masses ranging from $7.5 <$ \lmstar $< 9.1$ (median: $8.0$ \lmsol), and effective radii spanning $0.4 < (R_e/ {\rm kpc}) < 1.8$ (median: $1.0$ kpc). The extended sub-sample contains $21$ galaxies, with stellar masses between $7.9 <$ \lmstar $< 9.1$ (median: $8.5$ \lmsol), and sizes from $2.0 < (R_e/ {\rm kpc}) < 4.9$  (median: $3.0$ kpc). We note that classifying galaxies relative to a power-law of the mass-size relation yields a nearly identical grouping.

The top row of Figure \ref{fig:four_panel_compact_vs_extended} shows the compact sample represented by red points, while the extended sample is shown in blue in the second row. For both samples, we plot the values of all the galaxies at every snapshot ($z\leq3$). The first three panel columns show the same quantities as Figure \ref{fig:z0_tri_panel}, but we add an extra panel in both rows to show the temporal evolution of a single exemplary galaxy within their respective sample (right-most panels). We verified that short term outliers in size are attributed to the halo finder confusing merging bodies right before they coalescence. In the top size-sSFR panel (third panel of top row), we see all galaxies in the compact sample remain within the $\approx2$ kpc boundary while covering a wide range of sSFR since $z=3$. This radius is therefore adequate for encompassing and defining the compact galaxy sample. The panels in the last column show the size-sSFR evolutions of \galaxyname{r753} (compact) and \galaxyname{r642} (extended). \galaxyname{r642} begins as a compact galaxy similar to \galaxyname{r753}, but around $8$ Gyr into the simulation, transitions to a period of secular size growth (Section \ref{sub:Secular Size Growth}) that transforms it into an extended galaxy.       

\subsection{Secular Size Growth}
\label{sub:Secular Size Growth}

Figure \ref{fig:r_and_m_hist} shows the evolution of size and mass in our galaxy samples over cosmic time, with compact galaxies represented in red and extended galaxies in blue. Notably, the difference between compact and extended galaxies is not solely based on mass, as both groups display a diverse range of masses. Furthermore, for both samples, we do not observe significant fluctuations in size over cosmic time. As noted in the previous section, both extended and compact galaxies begin their evolution as compact galaxies with high sSFRs (as shown in Figure \ref{fig:four_panel_compact_vs_extended}). However, extended galaxies undergo secular growth in size, resulting in half-light radii that exceed $2$ kpc. The next section will explore the evolutionary paths that lead to the formation of extended galaxies.  

\section{Extended Galaxies and Stellar Disks}
\label{sec:Extended Galaxies and Stellar Disks}

Dwarf galaxies in our simulations are initially formed as compact structures but can transform into extended galaxies. A key detail associated with this kind of transformation is that all galaxies in our extended galaxy sub-sample have or recently had extended stellar disks at $z=0$. In this section, we investigate the relationship between stellar disks and extended galaxies. Specifically, we propose, for our stellar mass range, that the formation of stellar disks is the main pathway through which isolated dwarf galaxies evolve into extended galaxies. Furthermore, we find that these stellar disks form due to high-AM (spiraling-in) mergers.  

This section is divided into three main parts. Section \ref{sub:Disk Types and Examples} defines and categorizes disk galaxies according to the factors that influence their formation or disruption, providing individual galaxy examples for each scenario. Section \ref{sub:Disk Formation} provides details on how mergers influence morphology outcomes. Finally, Section \ref{sub:Conditions for Disk Growth} discusses the environmental conditions necessary for size growth via a stellar disk. 

\begin{figure*} 
\includegraphics[width=15cm]{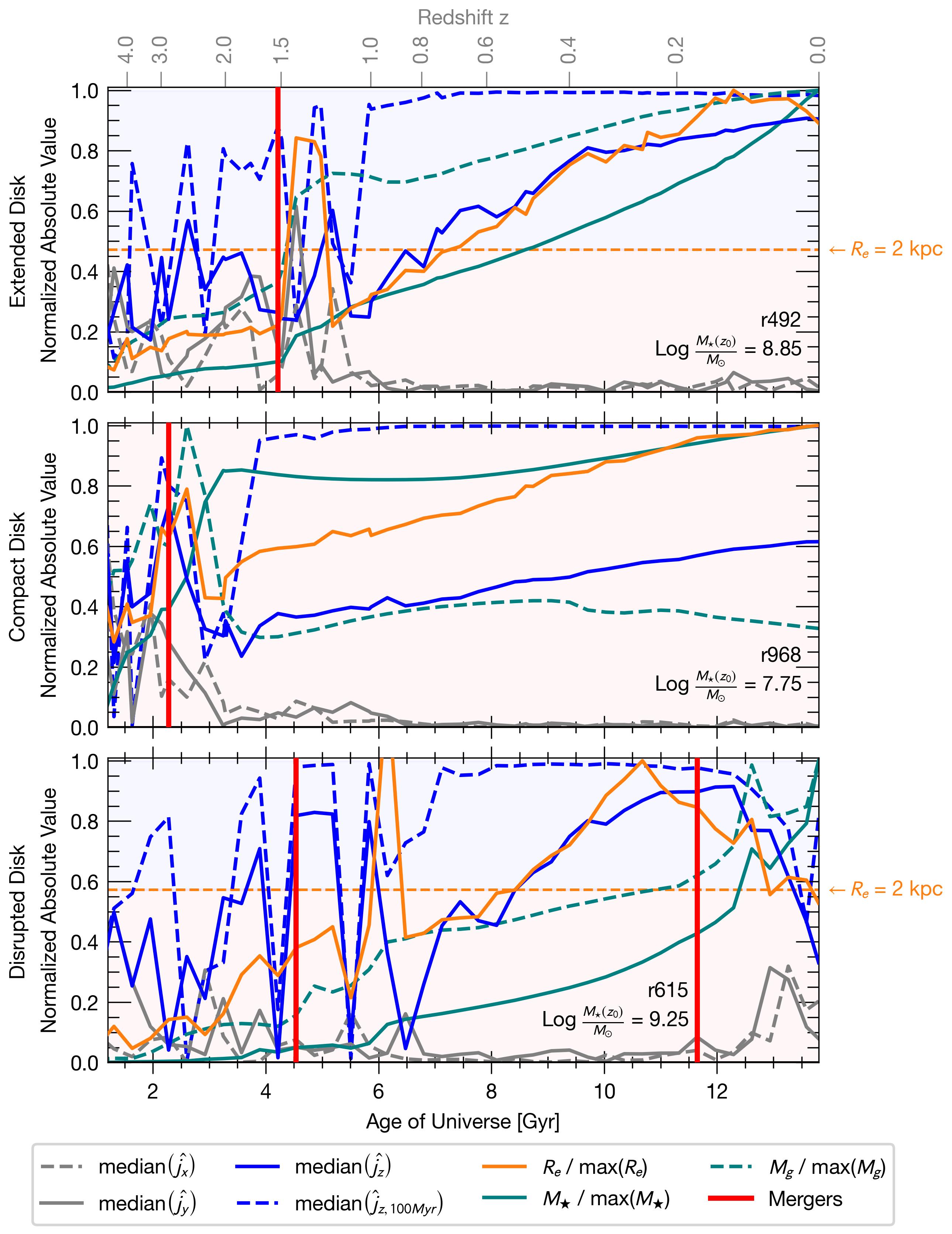}
\centering
\caption{\label{fig:single_j_plot}
Evolution in galaxy properties for \galaxyname{r492} (extended disk), \galaxyname{r968} (compact disk),  and \galaxyname{r615} (disrupted disk). We show the absolute value of the median $\hat{j_x}$ and $\hat{j_y}$ components in gray and $\hat{j_z}$ in solid blue for all the stars in the galaxy. We also include the absolute value of the median $\hat{j}_{z,100{\rm Myr}}$ for stars that formed within $100$ Myr of the snapshot as a dashed blue line. The half-light radius, normalized by the maximum value, is displayed in orange. To indicate the point at which the galaxy transitions into an extended galaxy, we include an orange horizontal line corresponding to $2$ kpc. The stellar mass and gas mass within the virial radius normalized to their maximum values are shown as solid and dashed teal lines, respectively. Lastly, the beginning of mergers that are responsible for morphological transformations are shown as red lines. 
}
\end{figure*}

\begin{figure} 
\includegraphics[width=8cm]{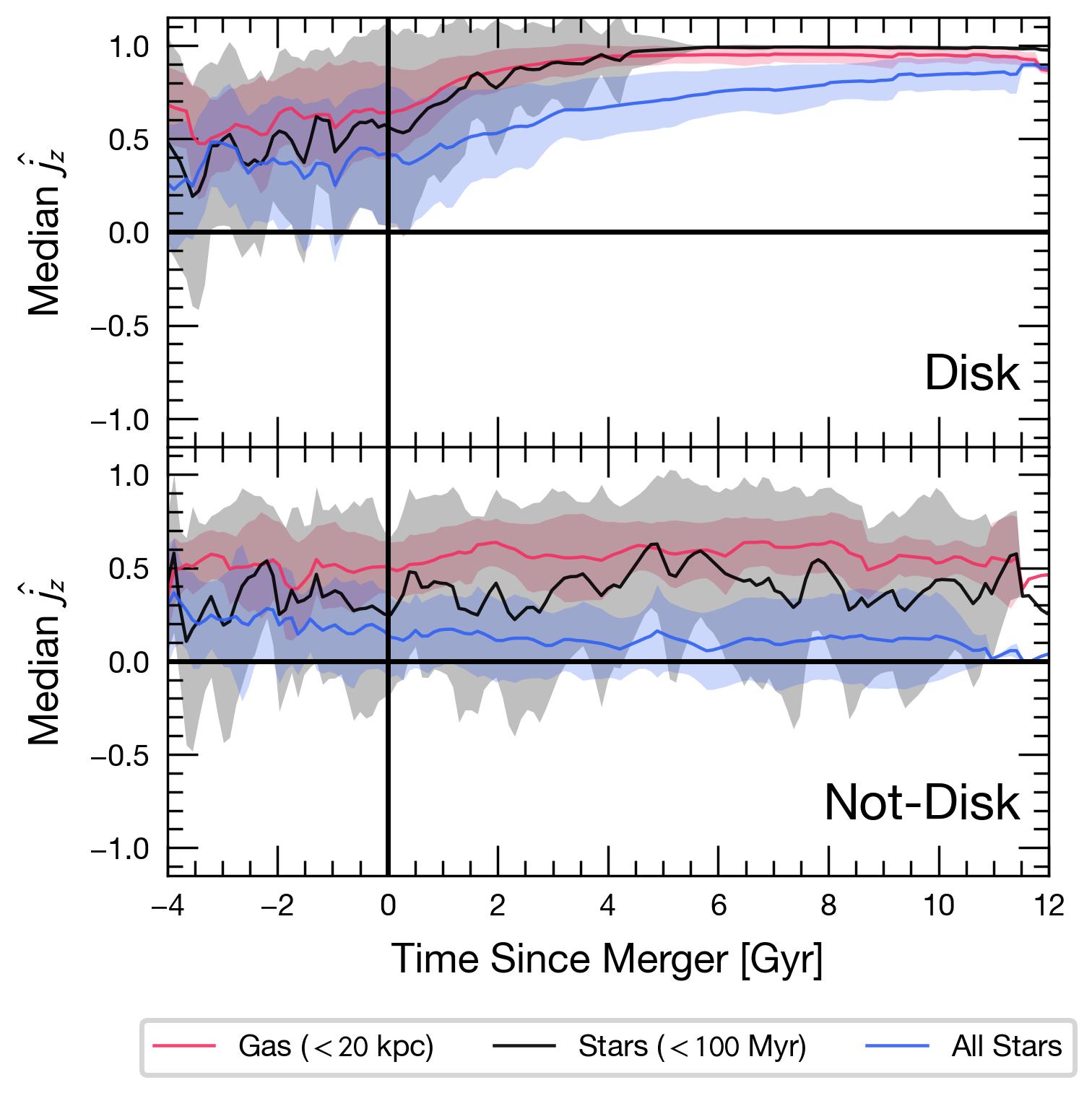}
\centering
\caption{\label{fig:time_since_merger}
Median $\hat{j_{z}}$ versus time since the highest AM merger for our disk (top) and not-disk (bottom) samples. Red, black, and blue curves show the gas within $20$ kpc, stars younger than $100$ Myr, and all stars, respectively. The solid line shows the mean values for all galaxies in the sub-sample at each timestep, while the shaded regions show the corresponding 1 standard deviation. For the disk sub-sample, the median $\hat{j_{z}}$ of the gas tends to $+1$ shortly after the mergers ($\lesssim 2$ Gyr) followed by new stars (star formation in the disk). As AM supported stars build up in the disk, the median $\hat{j_{z}}$ of all the stars is skewed towards $+1$. Not-disk galaxies have some rotational support but the median $\hat{j_{z}}$ of their stellar populations is near $0$ due to a lack of AM supported star formation.   
}
\end{figure}

\subsection{Disk Types and Examples}
\label{sub:Disk Types and Examples}

In this study, a disk galaxy is kinematically defined as one where the median $\hat{j_z}$ of its star particles exceeds $0.5$ (see Section \ref{sub:Disk Definition} for details). The subset of galaxies that satisfy this requirement are referred to as our ``disk sub-sample" and contains $25$ galaxies ($64\%$ of our final sample). Figure \ref{fig:disk_size_params} shows the size-sSFR and $j_\star$-size relations in the first two panels from the left. The last two panels show axis ratios $B/A$ and $C/A$ (where $A$ is the major axis, $C$ is the minor axis, and $B$ is the intermediate axis) measured by \cite{Keith2025} at $2R_e$ as a function of size.\footnote{Note that we do not use the same definition of ``disk'' as \cite{Keith2025}; they define disks based on axis ratios.} Our disk sample is colored in orange and our not-disk sample is in black. First, we see that all of the extended galaxies ($R_e > 2$ kpc), except for one, are disks at \zzero. Second, a key observation is that the disk galaxies have higher $j_\star$ magnitudes, by definition, since they are AM supported systems. Third, extended galaxies have relatively flatter stellar axis ratios than the compact galaxies ($0.2 \lessapprox C/A \lessapprox 0.35$, see \cite{Keith2025} for comparisons with \cite{2020ApJ...900..163K, 2021ApJ...920...72K}).  

The evolution of disk galaxies can be generalized into three evolutionary paths. Accordingly, we group the disk galaxies as extended disks (e.g., \galaxyname{r492}), compact disks (e.g., \galaxyname{r968}), and disrupted disks (e.g., \galaxyname{r615}). In this section, we present and discuss each evolutionary path by focusing on single galaxy cases that are typical or exemplary of their respective categories.

\subsubsection{Extended Disk Example}
\label{sub:Extended Disk Example}

\begin{figure*} 
\includegraphics[width=10cm]{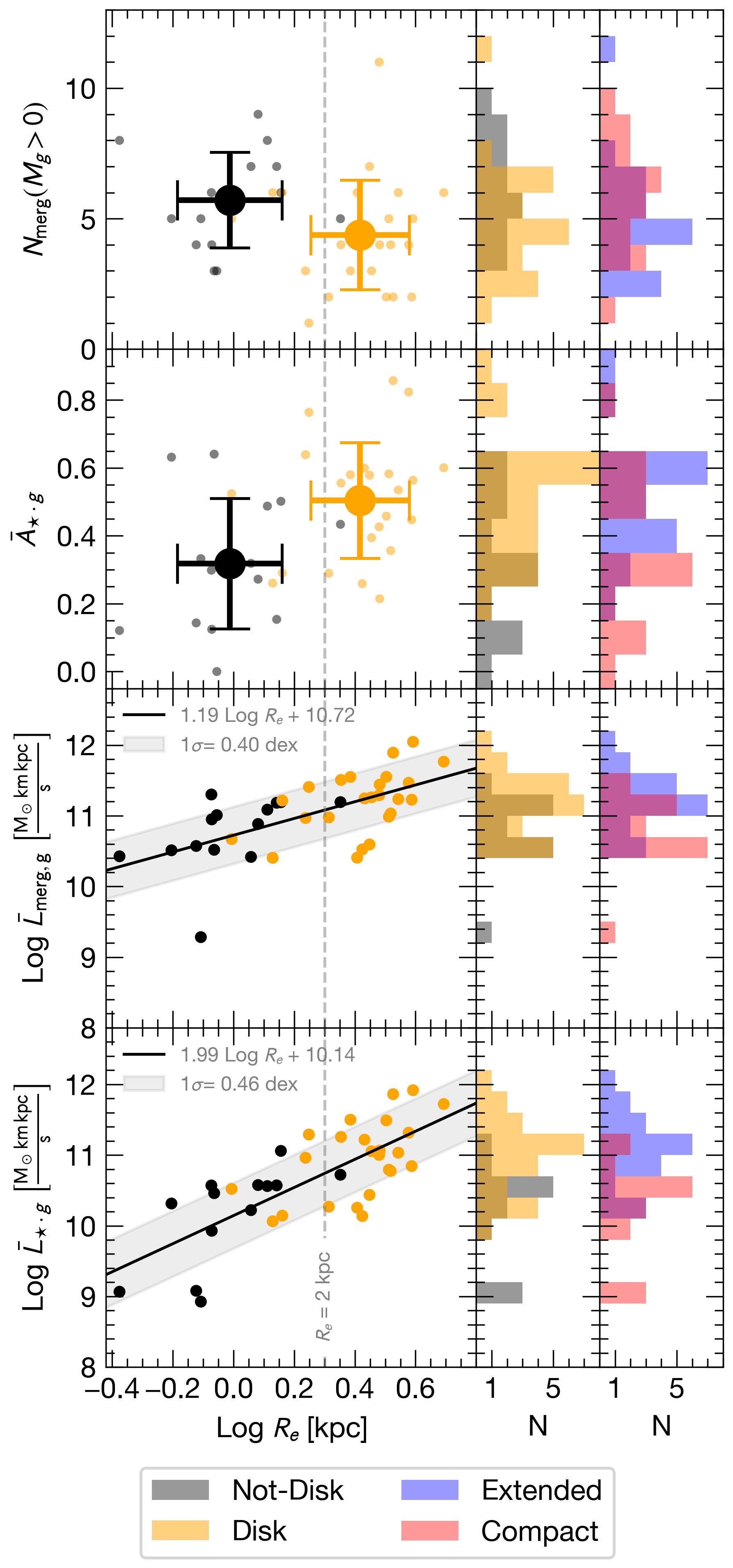}
\centering
\caption{\label{fig:Lsat_vs_r}
Mergers with gas containing satellites as a function of size ($R_e$). Top to bottom: The number of mergers with satellites that contain gas ($N_{\rm merg}(M_g > 0)$, i.e., galaxy with non-zero gas mass); The average alignment parameter of mergers ($\bar{A}_{\star \cdot g}$); The average AM of the gas in the merging satellites ($\bar L_{\rm merg,g}$); The average alignment-weighted AM of the gas in the merging satellites ($\bar L_{\star\cdot g}$). In the top two panels, we include the mean and one standard deviation for the disk and not-disk sub-samples as larger points with error bars. For each row, the right panels show histograms of the same quantity, colored by sub-sample. Though compact and extended galaxies span similar average AM, the misalignment of mergers in not-disk systems causes the alignment-weighted AM of compact galaxies to drop to lower values, resulting in a steeper slope.   
}
\end{figure*}

We explore a typical formation mechanism for extended disks by examining one of our representative extended galaxies, \galaxyname{r492}. This galaxy undergoes a merger with a gas-rich satellite (total mass ratio $1:2$ and gas mass ratio $1:3$) on a high-AM orbit (spiraling-in) relative to the primary. The merger starts (satellite enters virial radius) around $\sim4.2$ Gyr and coalesces at $\sim5.2$ Gyr. The gas from the merging satellite deposits AM into the gas surrounding the central galaxy, which reconfigures the gas into a disk (along the orbital plane of the satellite). Star formation starts in the newly formed gas disk, which creates stars with relatively high orbital AM. Over time, the gradual build-up of AM supported stars contributes to an increase in the half-light radius of the galaxy, eventually pushing its value beyond $2$ kpc (i.e. our extended galaxy definition). 

The top panel of Figure \ref{fig:single_j_plot} illustrates the sequence of disk formation and growth for \galaxyname{r492} by comparing several important normalized quantities as a function of time. In particular, we focus on trends in the normalized absolute values of each quantity. We plot the absolute value of the median $\hat{j_x}$ and $\hat{j_y}$ components in gray and  $\hat{j_z}$ in solid blue for all the stars in the galaxy.\footnote{There’s no need to normalize these values since they represent the distributions of unit vectors.} We also include the absolute median $\hat{j}_{z,100{\rm Myr}}$ for stars that formed within $100$ Myr of the snapshot as a dashed blue line. The half-light radius, normalized by the maximum value, is displayed in orange. To indicate the point at which the galaxy transitions into an extended galaxy, we include a dashed orange horizontal line corresponding to $2$ kpc. Finally, we represent the total gas mass within the virial radius with a dashed line and the total stellar mass with a solid line, both in teal. The snapshot corresponding to the onset of the merger is marked by the red vertical line.

In the top panel of Figure \ref{fig:single_j_plot} (\galaxyname{r492}), the galaxy starts off compact with the median $\hat{j_x}, \hat{j_y}$, $\hat{j_z}$, and $\hat{j}_{z,100{\rm Myr}}$ components fluctuating. After the merger, an initial deposit of gas that comes in with the merging satellite sharply increases the gas mass. The satellite deposits $\approx30\%$ of the gas mass within the virial radius at \zzero. Visual inspection of the merged body shows that the merger results in a gas disk around the central galaxy post-coalescence. The median $\hat{j}_{z,100{\rm Myr}}$ component spikes up to a value of $+1$ and remains there because the majority of the stars being formed are in the disk of the galaxy. Gradually, the accumulation of high-AM stars starts to skew the half-light radius as well as the median $\hat{j_z}$. The gas and star masses increase at a steady rate after the merger, indicating a relatively undisturbed and steady star formation history with ample gas supply. The steady star formation in the stellar disk is thus responsible for the secular increase in half-light radius for this kind of system.  

\subsubsection{Compact Disk Example}
\label{sub:Compact Disk Example}
In some cases, while a merger may result in a gas disk configuration, the feedback from the merger can disperse the gas if the galaxy is low mass. Though these galaxies are star forming post-merger, they lack ample gas reservoirs needed to accumulate sufficient AM supported stars to skew their $R_e$ (i.e., they stagnate in size). In this case, so long as the initial gas and stellar disks remain undisturbed by additional mergers, they will remain compact disks throughout cosmic time. 

To illustrate this formation channel, the middle panel of Figure \ref{fig:single_j_plot} shows an example of a compact disk galaxy, \galaxyname{r968}, with the same normalized values as the top panel. We omit the $(2\ {\rm kpc})/R_{e,{\rm max}}$ boundary in this figure because the galaxy's maximum half-light radius is only $1$ kpc. The formation of \galaxyname{r968}'s disk is initiated by a high-AM merger that begins $\approx 2.27$ Gyr into the simulation (total mass ratio $1:3$ and gas mass ratio $1:2$). This merger introduces AM into the gas surrounding the galaxy, leading to the creation of a disk aligned with the satellite's orbital plane. However, the merger also results in a spike in star formation, which generates substantial feedback, depleting the gas reservoir due to the galaxy's shallow potential. This depletion is evident in the plot, where we see a sharp decline in gas mass (dashed teal) at $\approx 3$ Gyr when the feedback occurs. 

The failure to produce a significant number of AM supported stars results in a slower growth of $R_e$ and median $\hat{j_z}$. It is also worth noting that the relatively weak star formation that does occur does so primarily within the compact disk. This is reflected by the high median value of $\hat{j}_{z,100{\rm Myr}}$ which persists throughout cosmic time. In contrast to our galaxy sample (all isolated), it is conceivable that in more crowded environments (clusters or groups), strong external influences either from nearby massive host galaxies or active galactic nuclei could deplete the gas supply of dwarf galaxies \citep[e.g.,][]{2006MNRAS.369.1021M, 2008ApJ...674..742B, 2022A&ARv..30....3B} shortly after they form their disks, rendering them compact disks \citep[e.g.,][]{2006AJ....132..497L, 2011A&A...526A.114T,2022AJ....164...18M}. 

\subsubsection{Disrupted Disk Example}
\label{sub:Disrupted Disk}
Dwarf galaxy disks are fragile structures that need a relatively undisturbed period to build up their stellar masses. There are many merger configurations that can destroy a disk, but in this section, we illustrate this evolutionary path by focusing on a single extreme example, \galaxyname{r615}. For this galaxy, an initial merger (enters virial radius at $\sim4.6$ Gyr, with total mass ratio $1:3$ and gas mass ratio $1:3$) triggers disk formation. Ample gas supply allows the galaxy to grow into an extended disk, with an evolutionary path similar to the example in Section \ref{sub:Extended Disk Example}. However, a second major merger (enters virial radius at $\sim11.7$ Gyr, with total mass ratio $1:3$ and gas mass ratio $1:3$) disperses the gas, as well as injects a non-negligible number of stars from the satellite. This disruption results in the redistribution of AM, and disturbed stellar orbits result in a corresponding decrease in half-light radius. We should note that this galaxy, which fully grows into an extended galaxy and then reverses into compactness due to a major merger, is an exception for our isolated galaxy sample.

\subsection{Disk Formation and Mergers}
\label{sub:Disk Formation}

\subsubsection{Mergers Make or Break}
\label{sub:Mergers Make or Break}

Stellar disks are structures that are supported by AM. Therefore, understanding their formation requires a look at the distribution and sources of angular momenta. As discussed in examples of Section \ref{sub:Disk Types and Examples}, high-AM satellite mergers play a critical role in creating gas disks that eventually form the \zzero stellar disk population. It is important to note that it is not the overall magnitude of AM injected that matters, since not-disk galaxies also experience and conserve high-AM interactions. It is instead the non-radial or spiraling-in orbit of the gas in the merging satellite that collisionally rearranges or ``stirs" the central galaxy’s gas into a disk configuration by applying dynamical torques and mixing (see Appendix \ref{appendix:Merger Example} for an example). Furthermore, the AM carried by the gas can be more important than the merger mass ratio itself \citep{2010MNRAS.408..783R}. This is because, for satellites on similar trajectories, a less massive satellite falling in from large distances may deliver as much AM as a more massive satellite that formed closer to the main galaxy. The AM of the satellite's gas is thus a good quantity to consider since it reflects how circular the orbit is, AM available for transfer, and gas mass available to drive disk formation. Therefore, the disk formation sequence can be summarized as follows:

\begin{enumerate}[leftmargin=1cm]
    \item A merging gaseous satellite on a non-radial (spiraling-in) orbit enters the virial radius of the main halo.
    \item  The satellite transfers its AM to the gas surrounding the central galaxy and redistributes the gas into a disk configuration through dynamical torques and mixing, usually along the plane of the satellite's orbit. Future smoothly accreted star forming gas inflows are primarily accreted onto this gas disk. 
    \item After coalescence, star formation begins and quickly dominates in the gas disk, resulting in an increasing population of AM supported stars.
    \item The accumulation of stellar mass in the disk leads to an increase in the half-light radius of the central galaxy.
\end{enumerate}

\begin{figure*} 
\includegraphics[width=\textwidth]{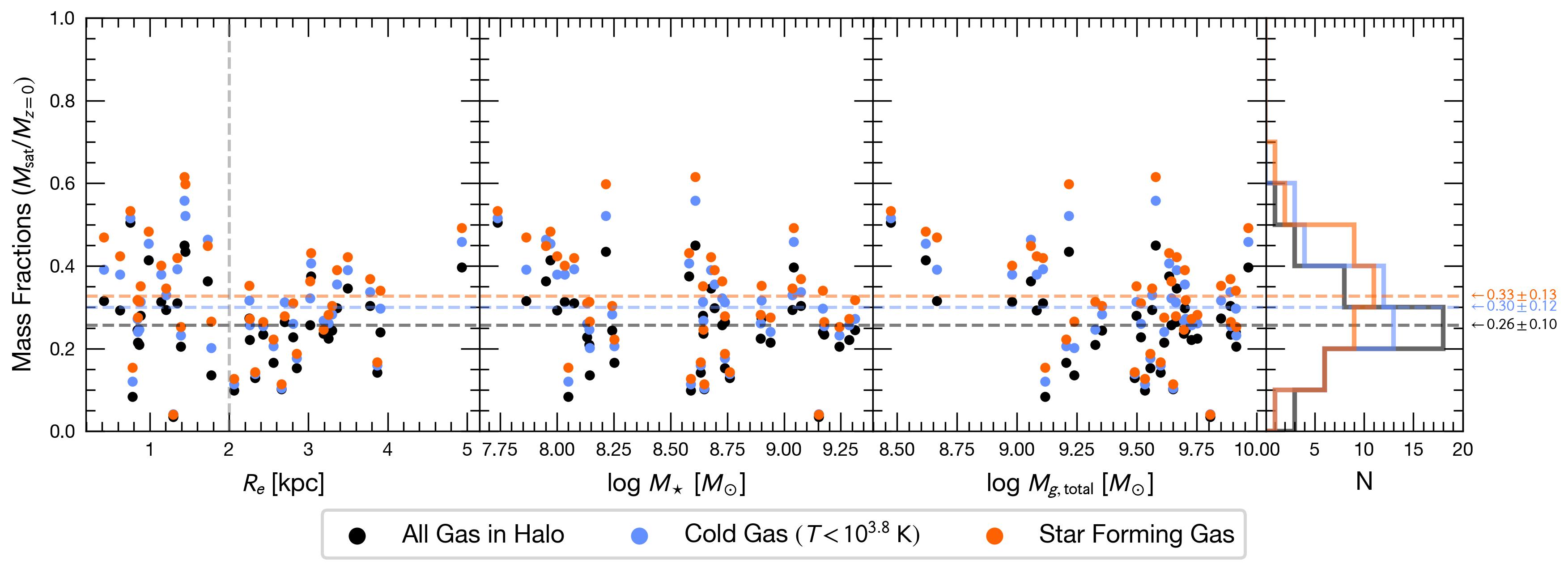}
\centering
\caption{\label{fig:gas_mass_fractions}
Mass fraction of $z=0$ cold (blue), star-forming (red), and total virialized gas (black) in the main galaxy that was previously accreted from merging satellites (since $z=6$). The panels show, from left to right, the satellite gas fraction as a function of galaxy size, stellar mass, and total gas mass, along with a histogram of the distribution. Mean contributions are $0.30 \pm 0.12$ for cold gas, $0.33 \pm 0.13$ for star-forming gas, and $0.26 \pm 0.10$ for all virialized gas. No strong trends are apparent, though there is a hint of increasing satellite contribution with galaxy size for extended systems.}
\end{figure*}

To illustrate this formation sequence the two panels of Figure \ref{fig:time_since_merger} present the median $\hat{j_{z}}$ for our disk and not-disk samples. The x-axis represents the time since the galaxy underwent its highest AM merger (usually the largest major merger). For each galaxy, we show the median $\hat{j_{z}}$ for the gas within $20$ kpc, new stars (age $<100$ Myr), and all the stars in the central galaxy using red, black, and blue colors, respectively.  

For disk galaxies, shortly after their merger, the median $\hat{j_{z}}$ for the gas rises to $+1$ (redistribution into a disk configuration), followed by a sharp increase in the median $\hat{j_{z}}$ for new stars. This pattern indicates that star formation shifts to primarily being driven by stars being formed in the gas disk. This transition typically occurs within approximately $2$ Gyr of the merger. The accumulation of stars in the disk gradually skews the median $\hat{j_{z}}$ of all the stars of the galaxy towards $+1$. We provide particle level details of the AM redistribution process using a single galaxy example in Appendix \ref{appendix:Merger Example}.  

In contrast, the bottom panel of Figure \ref{fig:time_since_merger} shows that the median $\hat{j_{z}}$ for new stars for not-disk galaxies fluctuates throughout cosmic time (the alignment is not consistent without a disk). It also shows that the AM of their gas, though it has a positive median $\hat{j_{z}}$, never gets dominated by its $z$ component for extended periods of time. Lastly, the wide distribution of median $\hat{j_{z}}$ for new stars causes the median $\hat{j_{z}}$ of all the stars to remain close to $0$ (i.e., there is no stellar build up with a preferred AM vector). The observation that the highest AM merger, which is often a major merger, does not change the distribution of AM like the disk galaxies suggests that the frequency, timing, gas fraction, and orientation of mergers are important factors.

It is important to note here that compact, extended, disk and not-disk galaxies are all experiencing mergers. The key factor that differentiates disk formation is the configuration of mergers. For dwarfs with disks, there are usually one or two dominating high-AM mergers that set the final distribution and orientation of the disk. If a disk galaxy (at \zzero) has undergone more than a single high-AM merger in its past, the orbital angular momenta of the merging satellites are usually co-rotating along the same plane.

In comparison to disk galaxies, not-disk galaxy formation mechanisms are complicated and diverse. One key observation for our sample of not-disk galaxies is that the mergers of these galaxies tend to be misaligned, counter-rotating, or head-on. For example, head-on collisions, which are often seen in our more massive not-disk galaxies, disperse gas widely, preventing the organized accumulation of gas and stars needed for a sustained disk. Another example is a galaxy that forms an initial disk but experiences a counter-rotating merger, which disrupts the gas disk or, if massive enough, disperses the stars in the stellar disk. Furthermore, 
strong feedback can play a role in gas disk disruption. In the next section, we explore how the alignment of mergers impacts the formation and disruption of stellar disks. 

\subsubsection{Merger Alignment}
\label{sub:Merger Alignment}

In this section we introduce a heuristic summary-statistic that links merger configurations to galaxy size. We use the angle between the satellite's orbital AM and the AM vector of the main galaxy's stellar disk as our alignment measure \citep[$\Delta\theta \equiv \measuredangle(
\vec{L}_{\rm main,disk}, \vec{L}_{\rm sat,orbit}
)$, similar to][]{NIHAO2019, 2021MNRAS.507.3301Z, Cadiou2022, Wu2025, Wright2025}. Building on this, we define our alignment parameter $A$ as the cosine of $\Delta\theta$, and limit its range from $[0,1]$:
\begin{equation}
    A \equiv
    \begin{cases} 
      0 & \Delta\theta > 90^{\circ} \\
      {\cos(\Delta\theta)} & \Delta\theta \leq 90^{\circ} \\
   \end{cases}, \label{eq:def_of_A}
\end{equation}
where $A=1$ represents maximum alignment and $A=0$ indicates head-on ($\Delta\theta = 90^{\circ}$) or counter-rotating ($\Delta\theta  > 90^{\circ}$) orbits. $A$ is calculated by taking the dot product of the two AM vectors. We use the AM vector of the stars at \zzero as a basis. This allows us to focus on the cumulative impact of the mergers and probe how closely the final stellar AM distribution is aligned with the orbital plane of the satellites. Furthermore the alignment parameter $A$, as defined in Equation \ref{eq:def_of_A}, does not attempt to distinguish or weigh if a head-on collision is more destructive than a counter-rotating merger. Because the AM transfer that is most important to gas disk formation is gas to gas interactions (see Appendix \ref{appendix:Merger Example} for example), we calculate the AM of the satellite gas before it merges. We calculate this value at the snapshot after in-fall where the satellite has the maximum orbital AM with respect to the \zzero orientation of the main galaxy's stellar disk. Thus, $A$ can be expressed as: 

\begin{align}
    A_{\star \cdot g} &= \max \left \{0, \cos \Delta\theta_{\star \cdot g} \right\} \\
    &= \max \left \{0,\,
    \frac{\vec{L}^{\,t=z_0}_{\rm main,\star}}{L^{\,t=z_0}_{\rm main,\star}}
    \cdot
    \frac{\vec{L}^{\,t={\max}(L_{\rm sat, g})}_{{\rm sat},g}}{L^{\,t={\max}(L_{\rm sat, g})}_{{\rm sat},g}} \right\}  , 
\end{align}

where $\vec{L}^{\,t={\max}(L_{\rm sat, g})}_{{\rm sat},g}$ is the AM vector of the gas in the merging satellite measured at $t={\max}(L_{\rm sat, g})$ and $\vec{L}^{\,t=z_0}_{\rm main,\star}$ is the AM vector of the main galaxy's stars at \zzero. $L^{\,t=z_0}_{\rm main,\star}$ and $L^{\,t={\max}(L_{\rm sat, g})}_{{\rm sat},g}$ are magnitudes of their respective vectors. For each central galaxy, we also compute the average AM of the gas in the merging satellites ($\bar{L}_{\rm merg,g}$), as well as the average alignment parameter ($\bar{A}_{\star \cdot g}$) and average alignment-weighted AM ($\bar{L}_{\star \cdot g}$) as follows:
\begin{equation}
    \bar{L}_{\rm merg,g} = \frac{1}{N_{\rm merg}}\sum_{\rm sat}^{N_{\rm merg}}   L^{\,t={\max}(L_{\rm sat, g})}_{{\rm sat},g}
\end{equation}
\begin{equation}
    \bar{A}_{\star \cdot g} = \frac{1}{N_{\rm merg}}\sum_{\rm sat}^{N_{\rm merg}}  A_{\star \cdot g} 
\end{equation}
\begin{equation}
\begin{split}
    \bar{L}_{\star \cdot g} & = \frac{1}{N_{\rm merg}}\sum_{\rm sat}^{N_{\rm merg}}   A_{\star \cdot g} 
    L^{\,t={\max}(L_{\rm sat, g})}_{{\rm sat},g} \label{eq:L_bar_a}
\end{split}
\end{equation}

where $N_{\rm merg}$ is the total number of merging satellites with gas and $\vec{L}^{\,t={\rm  max}(L_{\rm sat, g})}_{{\rm sat},g}$ is the maximum orbital AM of the gas of the satellite while it is within the virial radius of the main halo. We note that the alignment-weighted statistic, $\bar{L}_{\star \cdot g}$, is constructed to amplify alignment of mergers with the main galaxy's stellar AM at \zzero. More specifically, the restricted range of $A_{\star \cdot g}$ ($\Delta\theta_{\star \cdot g}<90^{\circ}$) means that we do not attempt to distinguish the relative destructiveness of counter-rotating interactions compared to direct head-on mergers. Instead, $A_{\star \cdot g}$ serves to indicate how much aligned gas AM contributes constructively to disk formation (i.e. constructive to gas disk formation). More importantly, these subtleties, which will be quantified at the end of subsequent paragraph, do not alter our conclusions.

Figure \ref{fig:Lsat_vs_r} shows the number of mergers ($N_{\rm merg}$), the average alignment ($\bar{A}_{\star \cdot g}$), $\bar{L}_{\rm merg,g}$, and $\bar{L}_{\star \cdot g}$ as a function of size for our sample of galaxies, with histograms to show distributions for our sub-samples. The first row shows that galaxies of all types experience similar numbers of mergers with satellites that contain gas, though disks and extended galaxies tend to have slightly fewer. Though the second row shows that disk galaxies tend to have slightly higher alignment parameter averages, it does not discriminate the most impactful mergers (i.e. mergers with high-AM). The third panel shows that $\bar{L}_{\rm merg,g}$ is similar between disk and not-disk populations, with extended galaxies having slightly higher values ($\log \bar{L}_{\rm merg,g} \propto 1.19\ \log R_e$). Lastly, the bottom panel shows $\bar{L}_{\star \cdot g}$, with misalignment causing the values of compact or not-disk objects to drop lower than the extended disk sample ($\log \bar{L}_{\star \cdot g} \propto 1.99\ \log R_e$). It is worth noting that an unrestricted alignment parameter (i.e. $A(\Delta\theta) \equiv \cos(\Delta\theta)$) or a linear function of $\Delta\theta$ (i.e. $A(\Delta\theta) \equiv \max\{0,\ 1-\Delta\theta / 90^{\circ}\}$) produce a similar slope ($\log \bar{L}_{\star \cdot g} \propto 1.99\ \log R_e$ and $\log \bar{L}_{\star \cdot g} \propto 2.10\ \log R_e$, respectively).

\subsubsection{Gas Contribution from Satellites}
\label{sub:Gas Contribution from Satellites}

\begin{figure*} 
\includegraphics[width=\textwidth]{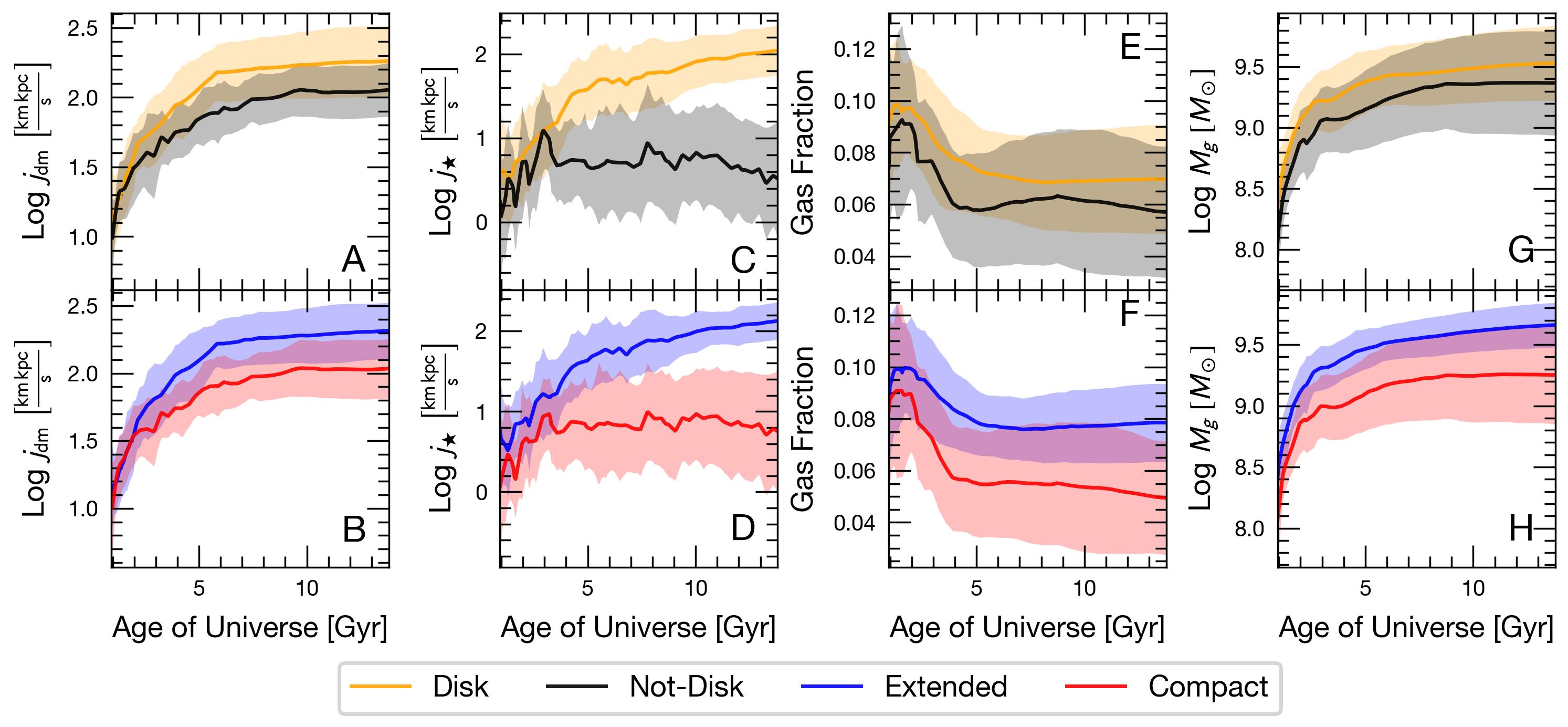}
\centering
\caption{\label{fig:pop_plots}
The four columns show the evolution of the specific AM magnitudes ($j_c \equiv |\vec{j_c}|$) of dark matter ($j_{\rm dm}$) and stars ($j_{\star}$), gas fraction, and total gas mass over cosmic time. In all panels the solid line shows the mean values for all galaxies in the sub-sample at each timestep, while the shaded regions show the corresponding 1 standard deviation. Both rows display the same galaxies: the top row shows disk systems (orange) and not-disk systems (black), while the bottom row shows compact systems (red) and extended systems (blue). The evolution of $j_{\rm dm}$ shows that there is no clear distinction between all sub-samples, though extended galaxies have slightly higher $j_{\rm dm}$ due to their larger masses (orbital velocities) and halo sizes. Both the disk and extended sub-samples show higher $j_{\star}$ values relative to the not-disk and compact sub-samples, respectively. This is because stellar disks are AM supported structures and because disks are the primary means by which compact galaxies grow into extended galaxies. Lastly, the gas fractions and masses show that disk and not-disk galaxies span similar ranges but that extended galaxies tend to have higher gas supplies than compact galaxies, which fuels the buildup of angular-momentum supported stars.
}
\end{figure*}

\begin{figure}[t] 
\includegraphics[width=8cm]{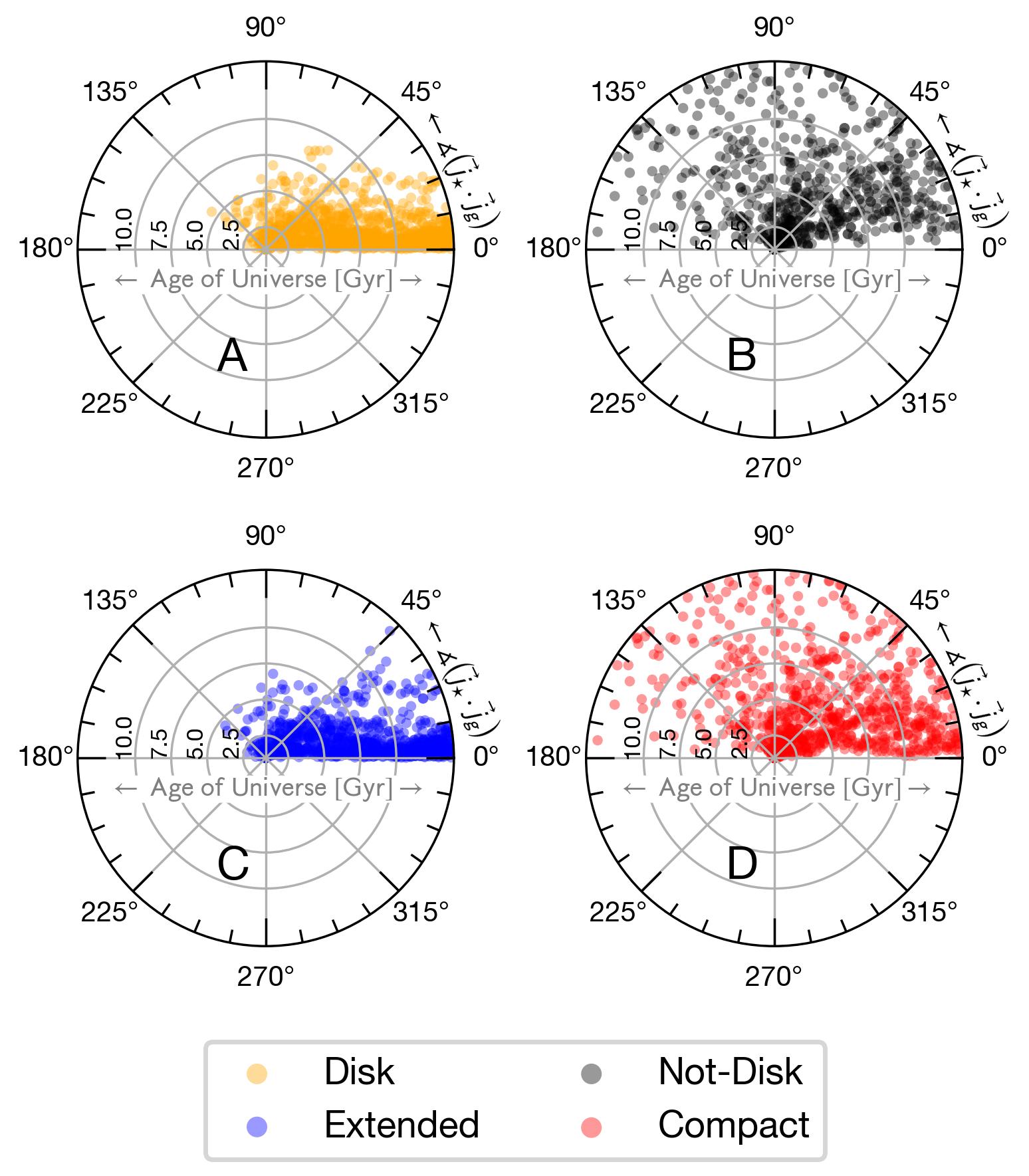}
\centering
\caption{\label{fig:pop_plots_polar}
Polar plots of the angular separation between gas and stellar AM vectors within $20$ kpc as a function of cosmic time (radial axis) and separation angle $\theta$ (azimuth) for (A) disk galaxies, (B) not-disk galaxies, (C) extended disks, and (D) compact galaxies. Disk samples exhibit strong alignment, while not-disk systems show random distributions.
}
\end{figure}

Having demonstrated the importance of AM  alignment in disk formation, we now investigate the extent to which gas from the satellite actually helps grow the primary disk. In particular, we consider the fraction of \zzero cold, star forming, and total gas that was once contained within a merging satellite.  We define cold gas as gas particles with temperatures below $10^{3.8}\ {\rm K}$ and star forming gas as the progenitor of any star particle in the main galaxy's halo at \zzero. Following \cite{Brooks2009}, we define and tag gas particles from the satellite as particles that have been members of the satellite halo before it enters the main halo's (or its progenitors') virial radius, at any snapshot. Since the flagging of satellite particles draws exclusively on halo catalogs (i.e., does not require merger trees), we consider snapshots up to $z=6$. Lastly, we define star-forming gas as gas particles that have produced stars, at any point, by \zzero.

We find the average contribution of cold gas from satellites at \zzero is $0.30 \pm 0.12$, while for star-forming gas it is $0.33 \pm 0.13$. For all virialized gas, the mean contribution is $0.26 \pm 0.10$. The panels of Figure \ref{fig:gas_mass_fractions} show, left to right, the fraction of gas from satellites in each gas category as a function of size, stellar mass, and total gas mass, as well as a histogram of the distribution of fractions. These values are set by the baryon fraction of merging satellites at this parent halo mass and we do not observe strong trends for these quantities. We note however that compact galaxies show more scatter (with higher fractions) and that there may be a trend of increasing fraction at larger $R_e$ for the extended galaxies considered alone ($R_e$ versus fraction).

Lastly, individual galaxy gas particle distributions (histograms) of distance (up to the virial radius), temperature, and $\hat{j_z}$ indicate that gas from satellites is mixed with all the gas within the main halos by \zzero. This is expected because gas from gas-rich mergers mixes with the central galaxy as it collisionally redistributes the central galaxy's gas (see Appendix \ref{appendix:Merger Example} for an example). Further investigation beyond the scope of this work is needed to quantify merger-driven gas mixing and the role feedback-driven outflows play in further mixing satellite gas with the gas at higher distances from the center of the main galaxy by \zzero.

\subsection{Conditions for Disk Growth}
\label{sub:Conditions for Disk Growth}

While high-AM mergers trigger the formation of an initial gas disk through AM transfer, the evolution of the disk into a fully extended structure relies on subsequent processes. In particular, the presence of a gas supply and a period of relative dynamical quiescence allow the initial disk to stabilize and grow in stellar size into an extended galaxy. Therefore two important conditions must be met for the disk to develop properly into an extended galaxy: 

\begin{enumerate}[leftmargin=1cm]
    \item There needs to be a gas supply surrounding the galaxy to fuel star formation in the disk (see Section \ref{sub:Gas Mass and Fractions} for discussion).  
    \item Because the disks of dwarf galaxies are sensitive to mergers, the disk needs a relatively undisturbed period to build up its stellar mass.  
\end{enumerate}

All extended galaxies fulfill these two conditions during secular growth into extended galaxies. If the first condition is not met, the galaxy will remain compact despite forming a disk. This situation is particularly possible in systems at the lower end of our galaxy sample mass. In these cases, the galaxy's weak gravitational potential is unable to counteract the feedback from the merger that forms the initial disk. As a result, much of the surrounding gas is expelled from the galaxy and new gas is not accreted later (see Section \ref{sub:Compact Disk Example} for an example). 

After a gas disk forms, dwarf galaxies also require relatively quiescent merger histories because subsequent major mergers can disrupt star forming gas disks. If a radial or misaligned merger is significant enough to destroy the gas disk but fails to form a new gas disk, future misaligned star formation (see Figure \ref{fig:time_since_merger}) can cause an irregular morphology and the galaxy to remain compact. Furthermore, if a galaxy with an extended stellar disk encounters a major disruptive merger at late times, the disruption can cause a shrinking in half-light radius since the AM that supports the extended disk is redistributed (see Section \ref{sub:Disrupted Disk} for an example). Interestingly, a transitory not-disk extended galaxy may be observed as the disrupted galaxy shrinks back into a relatively massive, not-disk, higher surface density object.

\section{Discussion}
\label{sec:Discussion}
This section discusses how the evolutionary processes explored in this work shape dwarf galaxy properties. Section \ref{sub:Trends in the Size-sSFR Relation} discusses overall trends in the size-sSFR relation. Section \ref{sub:Population-Level Trends} considers population trends in AM, gas fractions, and the alignment of stellar, gas, and halo components. Section \ref{sub:Disk Formation Comparisons} compares disk formation mechanisms across simulations and mass scales. 

\subsection{Trends in the Size-sSFR Relation}
\label{sub:Trends in the Size-sSFR Relation}

In Section \ref{sec:size-sSFR relation}, we present the size-sSFR relation of our sample and define two populations of galaxies based on size (a $R_e = 2$ kpc cutoff), which we label as compact and extended galaxies. In particular, we note that extended galaxies start as compact galaxies at high redshifts, but experience a period of gradual growth that increases their size beyond $2$ kpc. We find that the primary driver of size growth is the formation and gradual accumulation of stars in stellar disks. 

On timescales shorter than $1$ Gyr, we do not find appreciable changes in galaxy size in our simulations (see top panel of Figure \ref{fig:r_and_m_hist}). Pronounced size transitions in our sample, involving transitions between compact and extended regimes, are largely governed by the buildup or destruction of stellar disks. While we do not see size changes associated with SFR changes, we do see fluctuations in the SFR on $0.1-0.3$ Gyr timescales, in line with timescales as suggested by data \citep{2011ApJ...739....5W, 2019ApJ...881...71E,  2020MNRAS.498..430I, 2025arXiv250616510M}. A full exploration of that variation is beyond the scope of this work and will be addressed in a follow-up study. Differences in breathing modes (appreciable size fluctuation) between our sample and those predicted by \cite{ElBadry16} are potentially due to strongly bursty feedback in the FIRE simulations. Lastly, we note that our sample is purposefully restricted to isolated systems, which limits external perturbations that might otherwise trigger extreme starbursts or rapid quenching. As such, we do not capture the full range of specific star formation rates seen in denser environments, and in general, we expect isolated field dwarfs to remain star-forming \citep{2012ApJ...757...85G}.

As discussed in Section \ref{sub:Conditions for Disk Growth}, gas supply, which can be quantified by gas fractions, also plays a key role in shaping the size–sSFR relation. Disk galaxies with higher gas fractions can sustain higher star formation rates ($-10 \lessapprox {\rm sSFR}/{\rm yr}^{-1} \lessapprox -9$) and build extended stellar disks, carrying them into the extended and even low surface brightness regime \citep[e.g.,][]{NIHAO2019, Wright2025}. In contrast, systems with low gas fractions, usually due to feedback from mergers (see Section \ref{sub:Compact Disk Example} for gas poor disk galaxy), are more likely to exhibit suppressed star formation ($-11 \lessapprox {\rm sSFR}/{\rm yr}^{-1} \lessapprox -10$) and remain compact regardless of morphology. 

\subsection{Population-Level Trends}
\label{sub:Population-Level Trends}

In Section \ref{sec:Extended Galaxies and Stellar Disks} we discuss the evolutionary paths that lead to stellar disks and extended galaxies. In this section, we analyze trends at the population level that emerge from these evolutionary pathways. Specifically, we examine the evolution of AM alongside changes in gas fractions as a function of our sub-samples. We also consider the alignment of the stellar and gas components of each sub-sample. Figure \ref{fig:pop_plots} summarizes these trends: each column displays the magnitudes of $j_{\rm dm}$, $j_{\star}$, gas fraction ($M_g/M_{\rm total}$), and total gas mass within the virial radius ($M_g$) over time (Section \ref{sub:Angular Momenta}). Figure \ref{fig:pop_plots_polar} shows the evolving alignment between the AM vectors of gas (within $20$ kpc) and the stellar component (Section \ref{sub:Stellar and Gas Disk Alignment}). Disk systems are shown in orange, not-disks in black, compacts in red, and extended disks in blue.

\subsubsection{Angular Momenta}
\label{sub:Angular Momenta}
Panels A and B of Figure \ref{fig:pop_plots} show the mean $j_{\rm dm}$ of all the galaxies in each sub-sample as a function of cosmic time, while Panels C and D show the mean $j_{\star}$ (the shaded regions correspond to $\pm 1\sigma$). First, the specific AM of the dark matter distributions are overlapping at early times, with disk and extended galaxies having slightly higher mean $j_{\rm dm}$ over time. However, a very clear separation is evident in the mean $j_{\star}$ with disk and extended galaxies having significantly higher values. This can be understood as disks being supported by AM. The separation is also evident for extended vs compact galaxies because almost all of the extended sub-sample are disks.

\subsubsection{Gas Mass and Fractions}
\label{sub:Gas Mass and Fractions}
Panels E and F of Figure \ref{fig:pop_plots} show the mean gas mass fractions ($M_g/M_{\rm total}$) of each sub-sample as a function of time with $1\sigma$ shaded regions. The gas fractions span the same range for disk versus not-disk classifications but show a differentiation for compact versus extended sub-samples, with extended galaxies having higher mean gas fractions. The same trend can be seen in Panels G and H of Figure \ref{fig:pop_plots}, which similarly show the total gas masses. This trend arises because, after the formation of an initial disk, an ample supply of gas is necessary to facilitate the growth of the stellar population supported by AM. The build-up of AM supported stars forming in the AM supported gas increases the size of the galaxy into an extended galaxy (see Figure \ref{fig:r642_stars_post_merger} in Appendix \ref{appendix:Merger Example} for example).

\subsubsection{Stellar and Gas Disk Alignment}
\label{sub:Stellar and Gas Disk Alignment}
The polar plots of Figure \ref{fig:pop_plots_polar} show the angular separation between the gas (within $20$ kpc) and stellar $\vec{j}$ vectors (within the halo) as a function of time. The radial components represent cosmic time since $z=3$ and the $\theta$ coordinates show the angular separation. For the disk sample (Panel A), the gas and stellar angular momenta are largely aligned since star formation is predominantly contained within the gas disk. This aligns with the formation of extended galaxies from disks; thus, the extended sub-sample (Panel C) closely resembles the main population of disks. However, not-disk objects (B) reveal a more scattered distribution of gas and stellar alignments. Since the compact sub-sample (D) is dominated by not-disk galaxies, with a few compact disks, both scattered and some aligned distributions can be observed. 

These trends indicate that an important prediction regarding the evolutionary path of extended galaxies is that their stellar populations and gas disks are generally aligned over much of cosmic time. This alignment occurs because  the merger rate declines with time \citep[e.g.,][]{2010MNRAS.406.2267F}, so that most consequential disk-forming mergers are expected to have taken place at higher redshift. Afterward, star formation would have occurred predominantly within the plane of the gas disk. Accordingly, extended disk dwarf galaxies in nature are expected to host the majority of their younger stellar populations within their AM-supported stellar disks. 

\subsection{Disk Formation Comparisons}
\label{sub:Disk Formation Comparisons}

Many earlier studies have detailed the formation, evolution, and disruption of galactic disks in Milky Way mass systems \citep[e.g.,][]{1972ApJ...178..623T,2002MNRAS.333..481B,2005ApJ...622L...9S,2009ApJS..182..216K}. In particular, mergers have long been recognized as a key mechanism for disk formation and disruption \citep[e.g.,][]{1994ApJ...425L..13M, 1996ApJ...471..115B, 2002ApJ...581..799V, 2002MNRAS.329..423M, 2006ApJ...645..986R, Cox2006}. Subsequent studies demonstrated that the efficiency of disk regrowth depends on progenitor gas fractions and orbital parameters \citep[e.g.,][]{2006ApJ...645..986R,2009MNRAS.398..312G, 2005ApJ...622L...9S,2005MNRAS.361..776S,Cox2006,2009ApJ...691.1168H,2018MNRAS.473.2521S,2022MNRAS.516.5404S}, with high gas fraction encounters yielding disk dominated remnants. In this context, it is worth noting that angular momentum transfer remains central to whether disks survive or reform after mergers, and while our simulations benefit from high resolution, numerical effects in the central regions of low resolution simulations can bias sizes due to sensitivity of disk structure to AM loss \citep{2004ApJ...607..688G, 2007MNRAS.375...53K}.

In addition to AM content and gas fractions, recent cosmological simulations indicate that the orbit alignments of merging satellites can determine the size and morphology of the merger remnants of massive systems \citep{2008ApJ...683..597S,2009ApJ...691.1168H,2015ApJ...804L..40G,2018MNRAS.480.2266M, 2020MNRAS.494.5568J, 2021MNRAS.507.3301Z, Cadiou2022, Simons2024,Joshi2025}. \cite*{Cadiou2022} in particular show that, by modifying initial conditions (genetically modifying) the AM of regions in the simulation volume at $z\sim200$, the AM of stars in a halo can be causally linked to the AM of patches in the initial conditions. Specifically, changing the AM in initial conditions can predictably tune a galaxy’s stellar AM, half-light radius (up to $40\%$ increase), bulge fraction, and $v/\sigma$ at $z=2$. Most recently, \cite{Wu2025} found that, for galaxies with \lmstar$>10$ in the \texttt{IllustrisTNG} simulations \citep{2018MNRAS.475..676S, 2018MNRAS.480.5113M, 2018MNRAS.477.1206N, 2018MNRAS.475..648P, 2018MNRAS.475..624N}, spiral-in mergers yield remnants with larger disk but reduced bulge and hot inner halo fractions, whereas radial mergers show the reverse trend. Together, these works support the view that mergers and their orbital configurations are important factors for disk formation in massive systems. 

In addition to Milky Way mass galaxies, many have identified the processes by which simulations produce extended dwarfs through the lens of ultra-diffuse galaxy (UDG) formation \citep[see][for review]{2026PASA...43...31G}. Both FIRE \citep{2018MNRAS.478..906C} and NIHAO \citep{NIHAO2019} produce extended dwarfs through powerful gas outflows from stellar feedback that cause both the dark matter and the stars of dwarfs to expand. \cite{2019MNRAS.485..796M} and \cite{2021MNRAS.502.4262J} find that outflows also contribute to the formation of UDGs in HorizonAGN and the follow-up zoom-in simulation NewHorizon, although the larger driver is increased tidal perturbation at lower redshifts. In IllustrisTNG \citep{2023MNRAS.522.1033B,2024ApJ...977..169B} and Auriga \citep{2019MNRAS.490.5182L}, physically larger dwarfs inhabit higher spin dark matter halos \citep{2023MNRAS.518.5253Y}. \cite{2021MNRAS.502.5370W} also ties UDG formation in \texttt{Romulus25} (from which the initial conditions of the dwarfs studied in this paper are taken) to higher spin, but they find that the progenitors of diffuse dwarfs acquire this spin through high-AM major mergers and do not necessarily remain high spin at \zzero. This formation mechanism is explicitly linked to the formation of diskier dwarf galaxies in \texttt{Romulus25} in \cite{2022ApJ...926...92V} and mirrors the findings of \cite{NIHAO2019} and \cite{Wright2025} regarding the size evolution of slightly higher mass (\lmstar$>9.5$) galaxies in NIHAO and \texttt{Romulus25}, respectively. 

\subsubsection{Disk Formation and Concentration}
\begin{figure} 
\includegraphics[width=8cm]{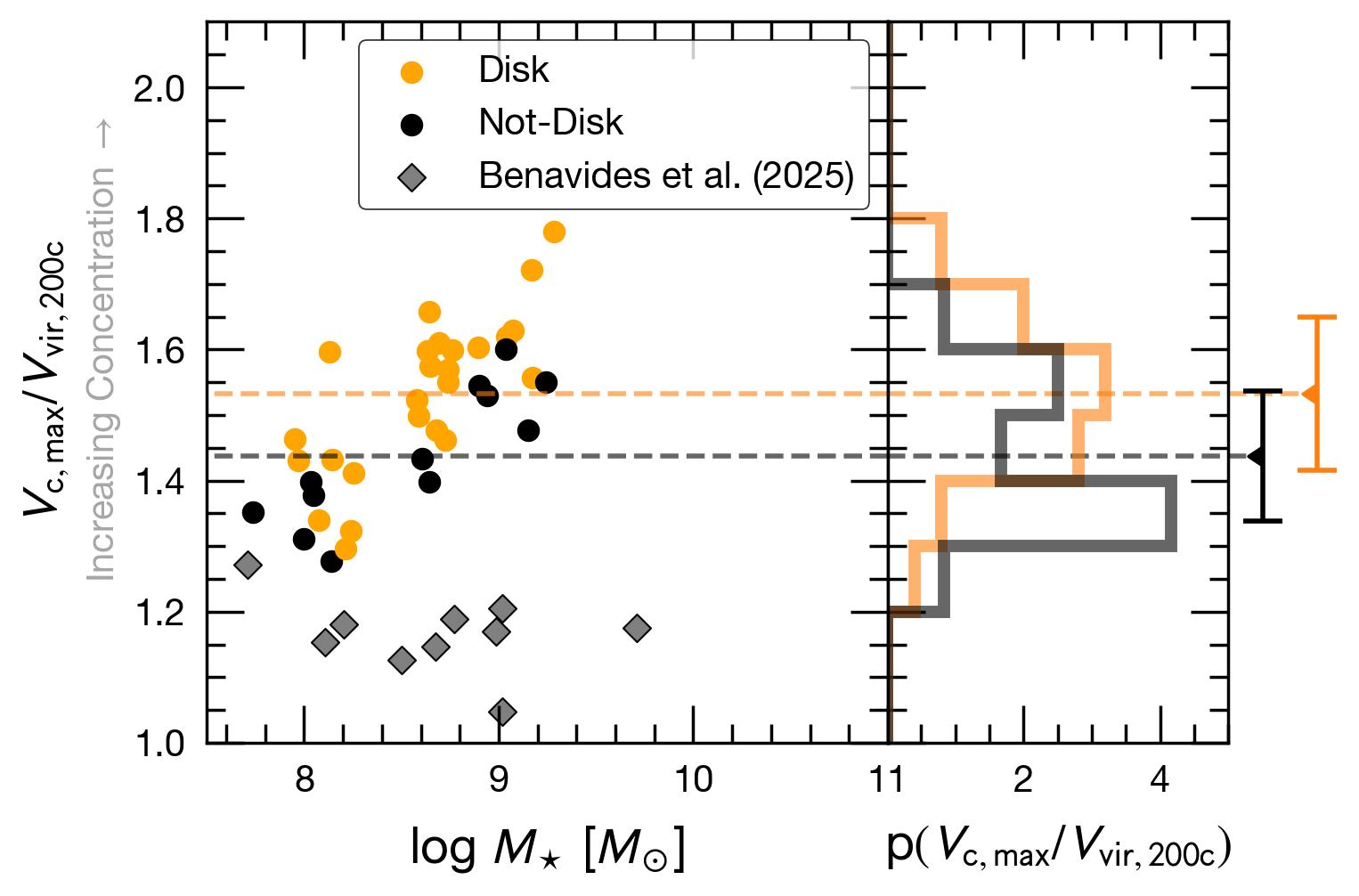}
\centering
\caption{\label{fig:concentration_hist}
$V_{\rm c, max} / V_{\rm vir, 200c}$, a proxy for halo concentration, as a function of stellar mass at \zzero, and its probability density histogram, for the disk and not-disk sub-samples. The probabilities are normalized by sub-sample. The disk sub-sample is in orange and the not-disk sub-sample is in black. The error bars at the right show that the two sub-samples are separated by $\sim1\sigma$ of their respective means. The gray diamonds show the $V_{\rm c, max} / V_{\rm vir, 200c}$ range of the zoom-in \texttt{FIRE-2} dwarf galaxies (\texttt{m11}'s, $7.6 <$ \lmstar $< 9.7$) in \cite{Benavides2025}, only one of which shows significant rotational support (see text). Both our disk and not-disk sub-samples are more concentrated than the \texttt{FIRE-2} dwarfs. Lastly, the Massive Dwarfs simulations produce both disk and not-disk object throughout the full mass range.
}
\end{figure}

\cite{Hopkins2023} use the \texttt{FIRE-2} simulations \citep{Hopkins2014,FIRE2} to study the growth of disks in dwarf galaxies specifically. They find that the key criterion for initially forming a disc is the development of a sufficiently centrally concentrated mass profile (e.g., ratio of $V_{\max}$ radius to the radius which contains $90\%$ of HI). The study finds a dwarf galaxy evolving from a bursty spheroidal system with a shallow potential to a system with a concentrated mass profile in a deep potential well, eventually forming a disk and exhibiting smooth star formation. In contrast, we find that disk formation in our sample is more strongly correlated with AM transfer from gaseous satellites. 

Most recently, \citet{Benavides2025} analyze a volume of \texttt{FIRE-2} and report a sharp mass–morphology trend: rotationally supported stellar disks are rare below \lmstar$< 9$, and become common only across a “transition” interval $9 <$ \lmstar $< 10$. The cold gas kinematics track the stellar morphology: in \texttt{FIRE} dwarfs the cold gas remains dispersion dominated, new stars inherit that state, and long-lived stellar disks do not form. \cite{Keith2025} find a lower transition mass for the Massive Dwarfs sample (\lmstar $\sim 8$). Lastly, observational studies constrain the transition mass, using the mass-size relation, to $8\lessapprox$ \lmstar $\lessapprox 9$ \citep{2011A&A...526A.114T, 2015ApJ...799..172T, Carlsten2021, 2021MNRAS.506..928N}.

\citet{Benavides2025} link late-time decline in star formation burstiness to the rise of disks, together with a deepening potential and the onset of a hotter CGM. We will examine the burstiness of our dwarf galaxies in a future work, but one possible reason why we are able to form disks at lower stellar masses might be that our feedback is weaker or less bursty \citep{Cruz2025}.  Since bursty feedback also leads to fluctuations of the gravitational potential and less steep potential wells (i.e., dark matter core creation), weaker feedback in our simulations might lead to deeper potential wells at lower masses and hence the ability to form disks at lower masses. Of particular interest for our results, the \galaxyname{m11b} zoom run in \citet{Benavides2025} shows a gas-rich, roughly coplanar minor merger at $z\sim1$ that delivers angular momentum, builds an extended thin gas disk, and produces young AM supported stars, an instance of disk growth by angular-momentum transfer from gaseous satellites that is consistent with the route we identify.  This is the only FIRE-2 zoom-in dwarf to show significant rotation, according to \citet{Benavides2025}.

Following \citet{Benavides2025}, we use the ratio $V_{\rm c, max} / V_{\rm vir, 200c}$ as a measure of concentration, where $V_{\rm c,max}$ is the maximum of the circular velocity profile ($V_{\rm c}(r) = \sqrt{G M(<r) / r}$) and $V_{\rm vir, 200c}$ is the virial velocity ($V_{\rm vir} = \sqrt{G M_{\rm vir,200c}/R_{vir,200c}}$). The values of $V_{\rm vir, 200c}$ and $V_{\rm c, max}$ are taken from the halo catalogs and rotation curves of these simulated galaxies as measured and described in \citet{Cruz2025}. Figure \ref{fig:concentration_hist} shows $V_{\rm c, max} / V_{\rm vir, 200c}$ as a function of mass and its probability distribution for our disk and not-disk sub-samples at \zzero (our disk definition is equivalent to using $\kappa$ down to $0.35$). The two sub-samples have similar concentration distributions, with disk galaxies being slightly more concentrated than the not-disk sample (a difference of order $1\sigma$). The gray diamonds in the figure mark the digitized values of $V_{\rm c,max}/V_{\rm vir,200c}$ measured for the \texttt{FIRE-2} dwarf galaxies (\texttt{m11} suite, $7.6 < \log M_\star < 9.7$) in \citet{Benavides2025}. Both of our sub-samples lie systematically above this range, indicating that even the not-disk galaxies in our sample are more concentrated than the typical \texttt{FIRE-2} dwarfs. Furthermore, there are not-disk galaxies in our sample which are as concentrated as disk galaxies, indicating that concentration alone may not be the criterion for disk formation though it is unclear if our disk formation scenario depends on concentration, as found in \citet{Hopkins2023}.  We will examine this further in future work. 

A related picture appears in \texttt{IllustrisTNG-50} (\texttt{TNG50}), but with different thresholds and mechanisms. \citet{Zeng2024} argue that many \texttt{TNG50} dwarfs form stars in disks, but subsequent episodes of star formation occur in gas disks misaligned with earlier ones, so the net stellar component becomes dispersion dominated. \citet{Celiz2025} took a closer look, finding that galaxies of all masses in \texttt{TNG50} form a dispersion-supported clump in their central $\sim$1 kpc.  This clump is likely numerical in origin, and in combination with the \texttt{TNG50} feedback methodology, they find that it prevents disk formation in the lowest mass galaxies.  \citet{Celiz2025} find a transition to disks shifted to lower masses,  $8 <$ \lmstar $< 9$. Lastly, \cite{2024A&A...687A.131D} compact dwarfs in \texttt{TNG50} form because they lack mergers that pump orbital angular momentum into the gas and enable size growth $z\sim0.8$ onwards. 

Taken together, FIRE and \texttt{TNG50} suggest that whether low-mass galaxies settle into long-lived disks depends on how feedback couples to the ISM, the coherence of late-time gas accretion, and the availability of orbital angular momentum from gas-rich companions \citep{Benavides2025,Celiz2025,Zeng2024}. \citet{Keith2025} have demonstrated that dwarfs in our simulated sample seem to become disk-dominated (defined using axis ratios) above \lmstar $\sim 8$, though they combined two sets of simulations with different feedback models, warranting a further study to ensure that feedback is not playing a role in the disk transition scale. We conclude by noting that the lowest-mass disk in our sample, \galaxyname{r997}, has a stellar mass of \lmstar $=7.7$, and the lowest mass extended galaxy, \galaxyname{r761}, has a stellar mass of \lmstar $=7.9$. An extended set of simulations with the same physics in \cite{Cruz2025} yielded a disk at \lmstar $=7.0$. A detailed study of the origin of these similarities and differences between simulation suites is left for future work.

\section{Summary and Conclusions} 
\label{sec:conclusion}
In this work, we explore the evolution of dwarf galaxies in the context of the size-sSFR relation, which provides a diagnostic of galaxy size while taking star-formation and feedback into account. We investigate the physical processes that influence the sizes of isolated dwarf galaxies using high-resolution cosmological zoom-in simulations of $39$ dwarf galaxies drawn from the ``Marvelous Massive Dwarfs" simulation suite ($7.5 <$ \lmstar $< 9.1$). The key findings from this work can be summarized as follows:

\begin{enumerate}
    \item We find dwarf galaxies with half-light radii ($R_e$) below $2$ kpc at \zzero have likely maintained sizes below $2$ kpc throughout their cosmic history (see Figure \ref{fig:four_panel_compact_vs_extended}). We label galaxies with $R_e < 2$ kpc as compact galaxies and $R_e \geq 2$ kpc as extended galaxies. 

    \item Though the mass–size relation empirically sets galaxy sizes on a global scale, extended and compact galaxies span our narrow stellar mass range and cannot be cleanly separated by stellar mass alone (see Figure \ref{fig:r_and_m_hist}).
    
    \item Extended galaxies exhibit gradual secular size growth over cosmic time while remaining star-forming at relatively stable sSFR. 

    \item The buildup of AM supported stellar populations in stellar disks is the primary pathway by which dwarf galaxies evolve into extended galaxies. The secular increase in size can be attributed to the formation and gradual buildup of stars in AM supported gas disks over cosmic time.
    
    \item Gas-rich mergers with high-AM (spiraling-in) and co-rotation redistribute AM in the central galaxy and create gas disks that eventually form stellar disks (see Appendix \ref{appendix:Merger Example} for an example). These mergers are thus the triggers that initiate the transition from compact to extended (see Figure \ref{fig:time_since_merger}).

    \item The growth of extended stellar disks requires not only the formation of a gas disk via high-AM mergers but also a sustained gas supply (see Section \ref{sub:Gas Mass and Fractions} for discussion) and a period of quiescence in the merger history of the galaxies as they build up their stellar population (see Section \ref{sub:Conditions for Disk Growth}).

    \item Some dwarf galaxies remain compact despite forming disks because feedback following the merger that created the disk can deplete their gas reservoirs, while in the lower mass regime, stronger numerical effects, such as spurious angular momentum loss, may also contribute to producing compact systems (see Section \ref{sub:Compact Disk Example} for an example). 
    
    \item Misaligned gas-rich mergers (e.g. counter-rotating or head-on) can disrupt pre-existing gas disks. Considering the cumulative effect, if we weight the AM contributed by merging satellites by its alignment with the stellar AM of the central galaxy at \zzero, we find that galaxies with well aligned satellite accretion tend to grow larger as their disks experience fewer disruptions (see Section \ref{sub:Merger Alignment} and Figure \ref{fig:Lsat_vs_r}).

    \item Considering all dwarf galaxies in our sample, on average, $30\%$ of the \zzero cold gas and $33\%$ of the star-forming gas originated from merging satellites since $z=6$, with no strong trends across galaxy properties such as size or mass (see Figure \ref{fig:gas_mass_fractions}).

    \item Population-level trends, such as stellar and gas disk alignment,  emerge from the evolutionary pathways discussed in this work. These trends are discussed in Section \ref{sub:Population-Level Trends}.
    \begin{itemize}
        \item  Disk galaxies show higher stellar specific AM magnitudes ($j_\star$) compared to not-disk galaxies (by definition, see Figure \ref{fig:disk_size_params}) but show similar ranges of gas fractions. 
        \item Because extended galaxies are mainly disk galaxies, extended galaxies have higher $j_\star$ than compact galaxies. Unlike disk versus not-disk galaxies, because ample gas supply is needed to grow in size, extended galaxies tend to have higher gas fractions than compact galaxies. 
        \item Disk galaxies have their gas and stellar angular momenta aligned at \zzero because star formation is dominated by the disk while not-disk galaxies show a large scatter in alignment. 
    \end{itemize}
\end{enumerate}

Looking ahead, a companion study will examine star formation histories of our sample in detail, followed by a test on how the morphologies described here depend on environment. We close by noting that several questions remain that require further investigation, such as the role and AM of inflowing gas from large-scale structure. Lastly, we look forward to future observational studies that span our sample's stellar mass range, such as the {\tt Merian} survey \citep{MERIAN2024, Danieli2024}, which will help to constrain the sSFR-size relation and its connection to morphology.

\section{Acknowledgments} 
We extend warm thanks to the authors of \cite{patel18} and \cite{Keith2025} for generously providing data used in the writing of this paper. We also thank Jacob Nibauer for useful comparative conversations about Milky Way-mass systems and the suggestion to use polar plots. We thank Daniel R. Piacitelli for helpful discussions on cooling times. We thank NSF/AAG grant 2506292 for providing funding support for this work. A.M.B acknowledges support from NASA Grant 80NSSC24K0894 and grant FI-CCA-Research-00011826 from the Simons Foundation. This work used Stampede2 at the Texas Advanced Computing Center (TACC) through allocation MCA94P018 from
the Advanced Cyberinfrastructure Coordination Ecosystem: Services $\&$ Support (ACCESS) program, which is supported by U.S. National Science Foundation grants 2138259, 2138286, 2138307,
2137603, and 2138296. Some of the simulations were performed using resources made available by the Flatiron Institute. The Flatiron Institute is a division of the Simons Foundation. This work was supported by NSERC (National Sciences and Engineering Research Council) of Canada. This research has made use of NASA's Astrophysics Data System. Software citation information aggregated using \texttt{\href{https://www.tomwagg.com/software-citation-station/}{The Software Citation Station}} \citep{software-citation-station-paper, software-citation-station-zenodo}.

\software{
\texttt{pynbody} \citep{pynbody}, \texttt{astropy} \citep{astropy:2013, astropy:2018, astropy:2022}, \texttt{matplotlib} \citep{Hunter:2007}, \texttt{numpy} \citep{numpy}, \texttt{python} \citep{python}, \texttt{scipy} \citep{2020SciPy-NMeth}, \texttt{ChaNGa} \citep{Jetley2008, Jetley2010, Menon2015}, and \texttt{Cython} \citep{cython:2011}}


\clearpage

\appendix
\restartappendixnumbering 

\section{Table of Halo Properties}
\label{appendix:Table of Halo Properties}

\begin{table}[H]
\centering
\noindent
\footnotesize
\begin{tabular}{l *{8}{c}}

\hline \hline
\multicolumn{1}{c}{Name} & \multicolumn{1}{c}{$M_{\rm total}$} & \multicolumn{1}{c}{$M_{\star}$} & \multicolumn{1}{c}{$R_{e}$} & \multicolumn{1}{c}{$R_{\rm vir}$} & \multicolumn{1}{c}{${\rm SFR}_{100 {\rm Myr}}$} & \multicolumn{1}{c}{${\rm sSFR}_{100 {\rm Myr}}$} & \multicolumn{1}{c}{Extended} & \multicolumn{1}{c}{Disk}\\ 
\multicolumn{1}{c}{(1)} & \multicolumn{1}{c}{(2)} & \multicolumn{1}{c}{(3)} & \multicolumn{1}{c}{(4)} & \multicolumn{1}{c}{(5)} & \multicolumn{1}{c}{(6)} & \multicolumn{1}{c}{(7)} & \multicolumn{1}{c}{(8)} & \multicolumn{1}{c}{(9)}\\ 
\hline
 & log $M_{\odot}$ & log $M_{\odot}$ & kpc & kpc & log $M_\odot$ yr-1 & log yr-1 &  & \\ 
\hline
r615$^\dagger$ & $11.22$ & $9.25$ & $1.85$ & $151.19$ & $0.10$ & $-9.15$ & \compactcolor{No} & \nodiskcolor{No}\\ 
r468 & $10.88$ & $9.09$ & $0.85$ & $113.23$ & $-0.51$ & $-9.61$ & \compactcolor{No} & \nodiskcolor{No}\\ 
r568 & $10.84$ & $9.06$ & $2.26$ & $109.81$ & $-0.45$ & $-9.52$ & \extendedcolor{Yes} & \diskcolor{Yes}\\ 
r488 & $10.92$ & $9.02$ & $1.39$ & $119.55$ & $-0.34$ & $-9.36$ & \compactcolor{No} & \nodiskcolor{No}\\ 
r442 & $10.92$ & $8.95$ & $2.43$ & $116.53$ & $-0.53$ & $-9.48$ & \extendedcolor{Yes} & \diskcolor{Yes}\\ 
r431 & $10.99$ & $8.95$ & $3.91$ & $123.13$ & $-0.62$ & $-9.57$ & \extendedcolor{Yes} & \diskcolor{Yes}\\ 
r502 & $10.90$ & $8.93$ & $1.29$ & $116.57$ & $-0.41$ & $-9.34$ & \compactcolor{No} & \nodiskcolor{No}\\ 
r492 & $10.90$ & $8.85$ & $3.78$ & $115.50$ & $-0.54$ & $-9.39$ & \extendedcolor{Yes} & \diskcolor{Yes}\\ 
r515 & $10.94$ & $8.82$ & $4.93$ & $118.51$ & $-0.63$ & $-9.45$ & \extendedcolor{Yes} & \diskcolor{Yes}\\ 
r556 & $10.77$ & $8.81$ & $1.21$ & $104.36$ & $-0.77$ & $-9.59$ & \compactcolor{No} & \nodiskcolor{No}\\ 
r555 & $10.79$ & $8.72$ & $0.85$ & $105.76$ & $-0.84$ & $-9.56$ & \compactcolor{No} & \nodiskcolor{No}\\ 
r489 & $10.86$ & $8.68$ & $2.25$ & $113.57$ & $-0.56$ & $-9.24$ & \extendedcolor{Yes} & \nodiskcolor{No}\\ 
r544 & $10.86$ & $8.68$ & $3.25$ & $111.01$ & $-0.63$ & $-9.30$ & \extendedcolor{Yes} & \diskcolor{Yes}\\ 
r656 & $10.71$ & $8.54$ & $2.32$ & $98.67$ & $-1.14$ & $-9.68$ & \extendedcolor{Yes} & \diskcolor{Yes}\\ 
r613 & $10.78$ & $8.52$ & $2.70$ & $105.68$ & $-1.08$ & $-9.60$ & \extendedcolor{Yes} & \diskcolor{Yes}\\ 
r597 & $10.76$ & $8.52$ & $2.85$ & $103.13$ & $-1.20$ & $-9.72$ & \extendedcolor{Yes} & \diskcolor{Yes}\\ 
r523 & $10.82$ & $8.51$ & $3.02$ & $107.60$ & $-1.17$ & $-9.68$ & \extendedcolor{Yes} & \diskcolor{Yes}\\ 
r642 & $10.73$ & $8.47$ & $3.36$ & $100.44$ & $-1.09$ & $-9.56$ & \extendedcolor{Yes} & \diskcolor{Yes}\\ 
r569 & $10.74$ & $8.46$ & $3.49$ & $102.25$ & $-1.20$ & $-9.66$ & \extendedcolor{Yes} & \diskcolor{Yes}\\ 
r918$^\dagger$ & $10.80$ & $8.45$ & $2.62$ & $106.36$ & $-1.35$ & $-9.80$ & \compactcolor{No} & \nodiskcolor{No}\\ 
r634 & $10.72$ & $8.43$ & $2.66$ & $101.06$ & $-1.20$ & $-9.63$ & \extendedcolor{Yes} & \diskcolor{Yes}\\ 
r571 & $10.81$ & $8.42$ & $3.19$ & $106.63$ & $-1.39$ & $-9.81$ & \extendedcolor{Yes} & \diskcolor{Yes}\\ 
r618 & $10.73$ & $8.42$ & $0.88$ & $101.06$ & $-1.21$ & $-9.63$ & \compactcolor{No} & \nodiskcolor{No}\\ 
r614 & $10.73$ & $8.41$ & $3.87$ & $100.72$ & $-1.33$ & $-9.74$ & \extendedcolor{Yes} & \diskcolor{Yes}\\ 
r563 & $10.73$ & $8.39$ & $1.43$ & $101.76$ & $-1.12$ & $-9.51$ & \compactcolor{No} & \nodiskcolor{No}\\ 
r886 & $10.65$ & $8.37$ & $2.06$ & $96.24$ & $-1.10$ & $-9.47$ & \extendedcolor{Yes} & \diskcolor{Yes}\\ 
r552 & $10.73$ & $8.36$ & $3.03$ & $100.24$ & $-1.24$ & $-9.60$ & \extendedcolor{Yes} & \diskcolor{Yes}\\ 
r741 & $10.59$ & $8.03$ & $2.56$ & $90.57$ & $-1.94$ & $-9.97$ & \extendedcolor{Yes} & \diskcolor{Yes}\\ 
r718 & $10.61$ & $8.02$ & $3.30$ & $94.48$ & $-2.05$ & $-10.07$ & \extendedcolor{Yes} & \diskcolor{Yes}\\ 
r719 & $10.57$ & $7.99$ & $1.45$ & $89.41$ & $-1.80$ & $-9.80$ & \compactcolor{No} & \diskcolor{Yes}\\ 
r749 & $10.61$ & $7.92$ & $1.77$ & $92.93$ & $-1.92$ & $-9.84$ & \compactcolor{No} & \diskcolor{Yes}\\ 
r716 & $10.59$ & $7.92$ & $0.86$ & $92.49$ & $-1.88$ & $-9.80$ & \compactcolor{No} & \nodiskcolor{No}\\ 
r761 & $10.59$ & $7.91$ & $2.81$ & $91.29$ & $-1.72$ & $-9.63$ & \extendedcolor{Yes} & \diskcolor{Yes}\\ 
r916 & $10.47$ & $7.85$ & $1.35$ & $82.41$ & $-1.85$ & $-9.71$ & \compactcolor{No} & \diskcolor{Yes}\\ 
r852 & $10.54$ & $7.83$ & $0.78$ & $86.73$ & $-2.01$ & $-9.84$ & \compactcolor{No} & \nodiskcolor{No}\\ 
r753 & $10.50$ & $7.81$ & $1.14$ & $83.77$ & $-2.32$ & $-10.13$ & \compactcolor{No} & \nodiskcolor{No}\\ 
r850 & $10.49$ & $7.78$ & $0.62$ & $84.50$ & $-1.99$ & $-9.76$ & \compactcolor{No} & \nodiskcolor{No}\\ 
r968 & $10.37$ & $7.75$ & $0.99$ & $76.45$ & $-2.52$ & $-10.27$ & \compactcolor{No} & \diskcolor{Yes}\\ 
r977 & $10.44$ & $7.73$ & $1.73$ & $81.42$ & $-2.53$ & $-10.26$ & \compactcolor{No} & \diskcolor{Yes}\\ 
r707 & $10.48$ & $7.64$ & $0.42$ & $83.09$ & $-2.73$ & $-10.37$ & \compactcolor{No} & \nodiskcolor{No}\\ 
r1023 & $10.19$ & $7.52$ & $0.75$ & $66.21$ & $-2.83$ & $-10.35$ & \compactcolor{No} & \nodiskcolor{No}\\ 
\hline 

\multicolumn{9}{l}{\footnotesize $\dagger$ Actively merging at redshift \zzero\ and excluded from final sample (r615 and r918)} \\

\end{tabular}
\caption{\label{table:overview}
Properties of the galaxy sample. Columns show: (1) galaxy name (simulation name - halo id at \zzero); (2) total mass $M_{\rm total}$; (3) stellar mass $M_{\star}$; (4) stellar half-light radius $R_{e}$; (5) virial radius $R_{\rm vir}$; (6) star formation rate averaged over 100 Myr ${\rm SFR}_{100,{\rm Myr}}$; (7) specific star formation rate over 100 Myr ${\rm sSFR}_{100,{\rm Myr}}$; (8) flag for extended stellar distribution; and (9) flag for presence of a disk. Galaxies excluded in the main analysis are marked with $\dagger$ after their name.}
\end{table}

\clearpage

\section{Particle Based Merger Example}
\label{appendix:Merger Example}

In the main text of this work, we explore the relationship between galaxy size and sSFR, as well as the process of disk formation using a range of galaxy-wide properties and statistics. In this appendix, we examine the particle distributions of a single galaxy to show how a high-AM merger drives disk formation and size growth. We specifically select a scenario where the central galaxy undergoes a single disk creating merger. This scenario simplifies the analysis, as it allows us to identify and track individual particles attributed to the satellite and directly assess the impact of the merger on redistributing the AM of the central. We chose \galaxyname{r642} because, in addition to fulfilling these criteria, its merger process takes a relatively longer time due to the high-AM of the satellite. The merger, which is a major merger ($1:3$), begins when virial radii of the main and satellite come into contact at cosmic time $t \approx 4.5$ Gyr and ends with coalescence at $t \approx 8.0$ Gyr.

We define our particle sample as all the gas particles in the main galaxy at \zzero. Similar to Section \ref{sub:Gas Contribution from Satellites}, we define and tag gas particles from the satellite as particles that have been members of the satellite halo before it enters the main halo's virial radius. We will refer to these gas particles as \textbf{satellite particles} and represent them in red. For the main halo, following \cite{Brooks2009}, we categorize the gas particles that end up in the main halo at \zzero into two groups: (1) \textbf{main particles}: These are particles that were part of the main halo prior to the merger at $t \approx 4.54$ Gyr. We represent these particles in blue. (2) \textbf{smooth particles}: These are particles that have become virialized as a result of smooth accretion, having never been a member of a satellite, after $t \approx 4.54$ Gyr. We represent these particles in green. The fraction of gas in each category is $0.33$, $0.16$, and $0.49$ for main, satellite, and smooth, respectively.  

\begin{figure*}
\includegraphics[width=14cm]{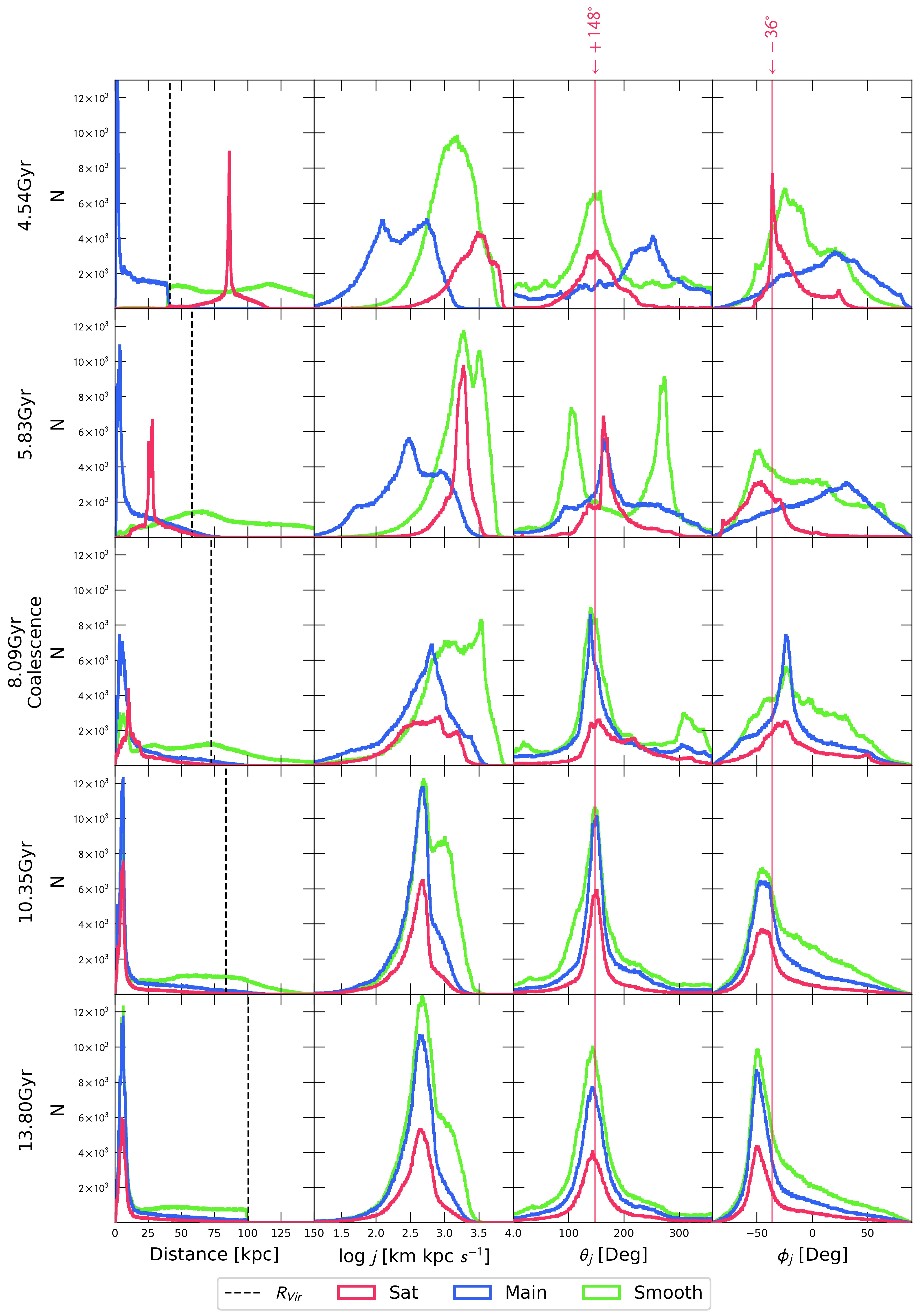}
\centering
\caption{\label{fig:r642_merger_all_gas}
AM distributions of \zzero gas particles of the main galaxy as a function of time. Rows correspond to different snapshots (time increasing downward). The four columns show, from left to right: distance from the galaxy center, and the polar coordinates of the particles' AM vectors relative to the center of the main galaxy ($\log j$, $\theta_j$, and $\phi_j$). Gas particles are divided into three categories: satellite particles (red), which were members of the satellite halo before in-fall; main particles (blue), which belonged to the main halo before the merger at $4.54$ Gyr; and smooth particles (green), which were accreted smoothly after the satellite enters the main halo. Red lines in the $\theta_j$ and $\phi_j$ columns indicate the orientation of the satellite’s gas orbital plane at $4.54$ Gyr. The histograms show AM injection and the satellite orienting the newly formed gas disk close to the plane of its orbit. The \zzero histograms show that particles from the main and satellite halos are mixed in phase space post-coalescence.    
}
\end{figure*}

First, we consider the AM distributions of the gas particles relative to the center of the main galaxy as a function of time. We use polar coordinates of each gas particle's AM vector because it allows us to separately analyze the injection of AM (the magnitude component, $\log j$) and how the merger sets the orientation of the disk (the $\theta_j$ and $\phi_j$ components). In this appendix, for each snapshot we center the volume to the center of the snapshot's main galaxy but maintain the original simulation orientation throughout all snapshots. Specifically, we do not rotate the coordinates to orient the main galaxy to a face-on configuration. Figure \ref{fig:r642_merger_all_gas} shows histograms of the number of particles for each gas membership category in each panel. The rows of the figure show different snapshots (time going down), while columns show, from left to right, distance from the main galaxy, $\log j$, $\theta_j$, and $\phi_j$. To indicate the orientation of the satellite's gas orbital plane, we add red lines in the $\theta_j$ and $\phi_j$ columns. The orbital plane's orientation is calculated by summing the AM vectors of the satellite's gas at $4.54$ Gyr ($(\theta_j,\ \phi_j) = (148^{\circ}, \ -36^{\circ})$). 

\begin{figure*} 
\includegraphics[width=14cm]{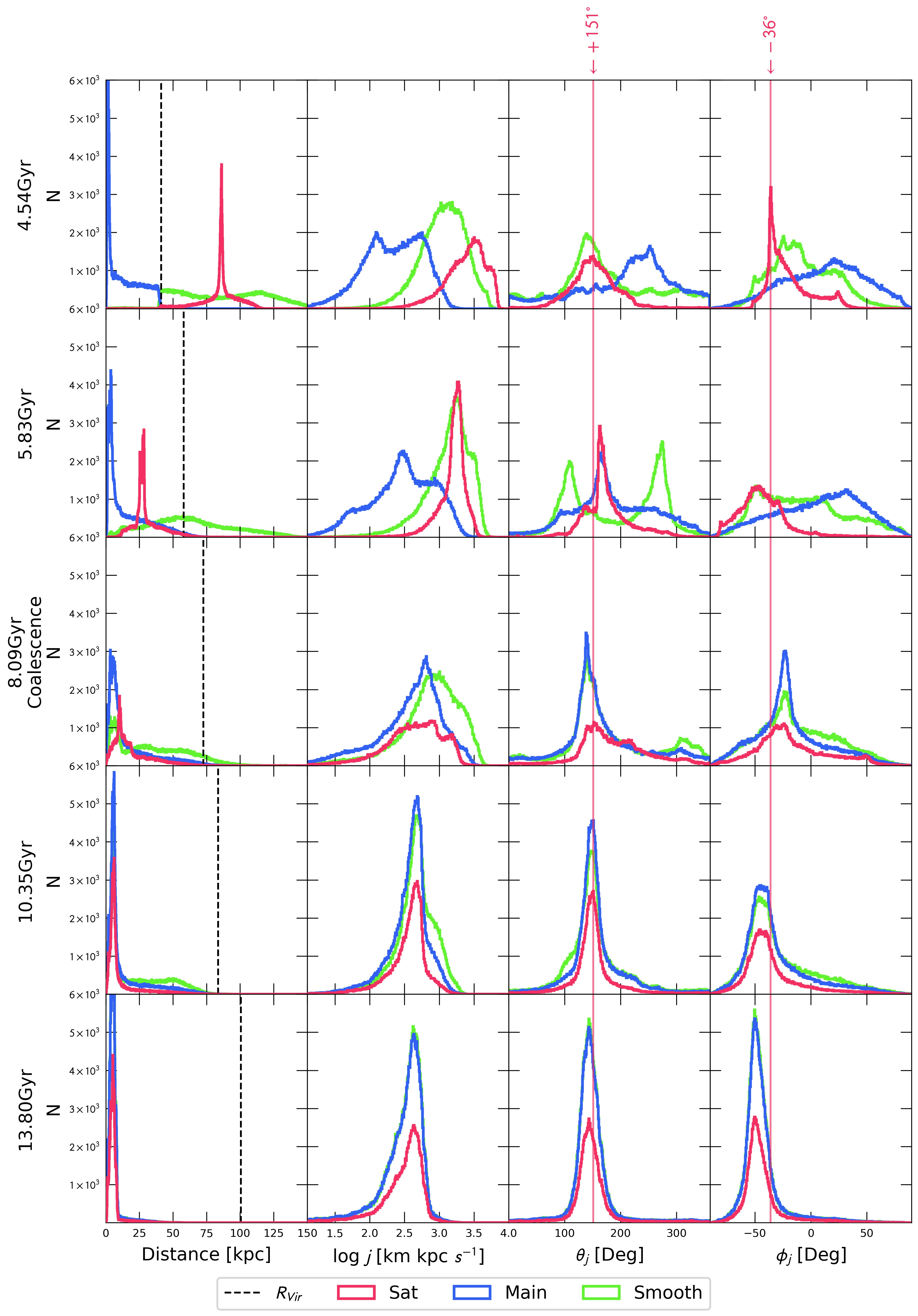}
\centering
\caption{\label{fig:r642_merger_cold_gas}
Same as Figure \ref{fig:r642_merger_all_gas}, but showing only \zzero cold gas particles ($T<10^{3.8}\ {\rm K}$) of the main galaxy as a function of time. The evolution of the histograms shows the same processes of AM redistribution as Figure \ref{fig:r642_merger_all_gas}. Furthermore, the three categories (main, sat, and smooth) of cold particles occupy very similar distributions in phase space and AM distribution by \zzero.  
}
\end{figure*}

At $4.54$ Gyr, the main galaxy has irregular morphology with some AM from an early merger (seen as a hump in $\theta_j$ and $\phi_j$). By virtue of its distance and orbital velocity, the satellite contains much larger AM than the main galaxy's gas at this point. At $5.83$ Gyr, the gravitational tidal and gas-to-gas torques begin to reorient and transfer AM to the main galaxy. Around $8.09$ Gyr, the two bodies coalesce, with the gases mixing spatially shortly after. At coalescence, the orientation and magnitude distributions of the angular momenta begin to align rapidly in magnitude and direction as hydrodynamic torques start to play a significant role in transferring AM. By $10.35$ Gyr, the gas that was in the main halo before the merger is redistributed to contain a larger amount of AM (transferred from the satellite), consolidated to a narrower distribution (the new gas disk), and oriented close to the orbital plane of the satellite. Figure \ref{fig:r642_merger_cold_gas} shows the same plots as Figure \ref{fig:r642_merger_all_gas}, but for a particle sample of all the cold gas particles in the main galaxy at \zzero. The evolution of the distributions shows the same processes of AM redistribution as all the gas. The fraction of cold gas in each category is $0.40$, $0.20$, and $0.40$ for main, satellite, and smooth, respectively.

\begin{figure*} 
\includegraphics[width=\textwidth]{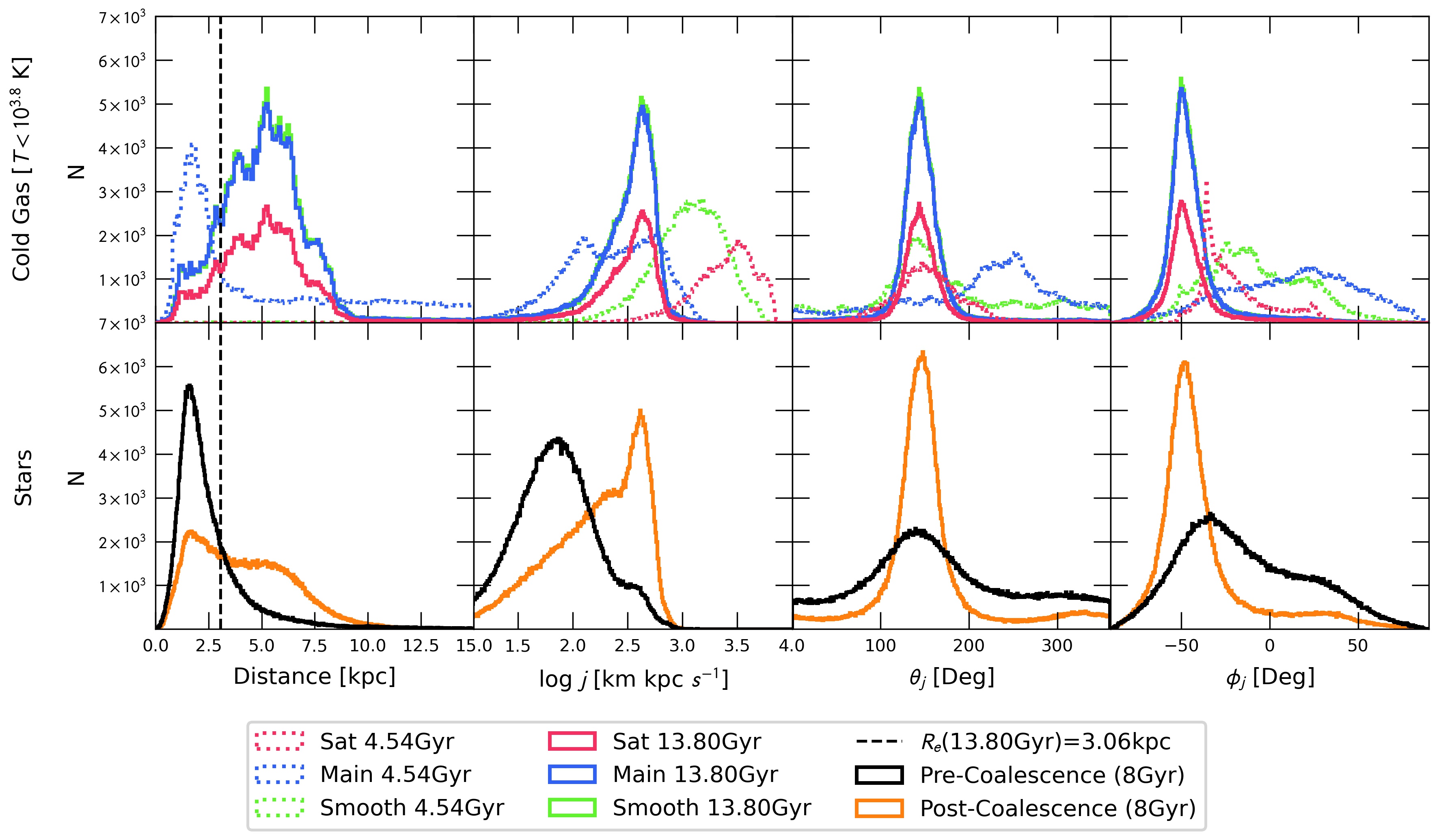}
\centering
\caption{\label{fig:r642_stars_post_merger}
Distributions of cold gas and stars in the main galaxy before and after the merger. The top row shows cold gas ($T<10^{3.8}\ {\rm K}$) distributions for the same quantities as Figures \ref{fig:r642_merger_all_gas}. The bottom row shows distributions for all stars at \zzero, split by formation time: stars formed before coalescence ($\approx 8.0$ Gyr; black) and after coalescence (orange). Distance distributions are shown within $15$ kpc, with the effective radius $R_e$ at \zzero marked (note that satellite and the smooth gas at $t=4.54$ Gyr are beyond $15$ kpc). The merger transforms the initially compact cold gas reservoir into an extended, AM supported disk, leading to the formation of a younger stellar population that inherits the gas disk’s AM distribution and orientation; this process builds up stars at larger radii and increases the effective size of the galaxy beyond $2$ kpc.
}
\end{figure*}

The top row of Figure \ref{fig:r642_stars_post_merger} shows the before and after merger distribution of the cold gas for the same quantities as Figure \ref{fig:r642_merger_all_gas} and \ref{fig:r642_merger_cold_gas}. The second row shows the distribution for all the stars in the main halo at \zzero, categorized by formation time. Stars that formed before coalescence ($\approx 8.0$ Gyr) are in black, and stars that formed after are in orange. Because the distance distribution is now zoomed in to the inner $15$ kpc we show the $R_e$ at \zzero instead of the virial radius. The gas distributions before the merger show that most of the main galaxy's gas was concentrated within $2.5$ kpc, with a majority of the pre-coalescence star particles occupying a similar distribution. The merger causes the gas to spread out by creating AM-supported gas, as discussed before. As the post-coalescence AM supported cold gas forms stars, it causes a build up of stars at higher radii (seen as a second hump in the post-coalescence stars). This buildup of AM supported stars increases the stellar size of the galaxy, driving it beyond $2$ kpc. Lastly, the post-coalescence stellar population inherits the AM distribution and orientation of the cold gas disk.

\clearpage
\bibliography{main}{}
\bibliographystyle{aasjournal}

\end{document}